\documentclass[aps,tightenlines,superscriptaddress,nofootinbib,showpacs,twocolumn]{revtex4}
\usepackage{amsmath, amsthm, amscd, amssymb, mathtools,graphicx}
\usepackage{bbm}
\usepackage{verbatim}

\begin{document}

\title{
Spontaneous collapse: a solution to the measurement problem and a source of the decay in mesonic systems}

\author{Kyrylo Simonov}
\email{Kyrylo.Simonov@univie.ac.at}
\affiliation{Faculty of Physics, University of Vienna, Boltzmanngasse 5, 1090 Vienna, Austria}
\author{Beatrix C. Hiesmayr}
\email{Beatrix.Hiesmayr@univie.ac.at}
\affiliation{Faculty of Physics, University of Vienna, Boltzmanngasse 5, 1090 Vienna, Austria}


\begin{abstract} Dynamical reduction models propose a solution to the measurement problem in quantum mechanics: the collapse of the wave function becomes a physical process. We compute the predictions to decaying and flavor--oscillating neutral mesons for the two most promising collapse models, the QMUPL (Quantum Mechanics with Universal Position Localization) model and the mass-proportional CSL (Continuous Spontaneous Localization) model. Our results are showing (i) a strong sensitivity to the very assumptions of the noise field underlying those two collapse models and (ii) under particular assumptions the CSL case allows even to recover the decay dynamics. This in turn allows to predict the effective collapse rates solely based on the measured values for the oscillation (mass differences) and the measured values of the decay constants. The four types of neutral mesons ($K$-meson, $D$-meson, $B_d$-meson, $B_s$-meson) lead surprisingly to ranges comparable to those put forward by Adler (2007) and Ghirardi-Rimini-Weber (1986). Our results show that these systems at high energies are very sensitive to possible modifications of the standard quantum theory making them a very powerful laboratory to rule out certain collapse scenarios and studying the detailed physical processes solving the measurement problem.
\end{abstract}

\maketitle

\section{Introduction}\label{sec:Introduction}
Quantum mechanics has proven to be an exceedingly successful theory. It covers a plethora of physical phenomena at different energy scales and, up to date, no experiments are in contradiction. However, quantum theory is very counter-intuitive and meets conceptual problems. Considering quantum mechanics as a fundamental theory also superpositions of macroscopic objects, such as cats, should exist which are obviously not observed in our daily world. In the Copenhagen interpretation during a measurement process a breaking of the superposition is mathematically postulated, but no detailed physical process has been assigned to it. Moreover, separation into macroscopic system (measurement apparatus) and microscopic system (quantum system) is utilized but lacks a clear definition. Ruling out unobserved macroscopic superpositions is the heart of the so-called measurement problem or macro-objectification problem~\cite{BassiGhirardi2003, BassiLochanSatin2013}. Dynamical reduction models, so-called collapse models, introduce an ontologically objective mechanism of the wave-function collapse. Since collapse models provide definite predictions for the regime between microscopic and macroscopic they are experimentally testable.

In 1985 the first dynamical reduction model, the GRW model, proposed by Ghirardi, Rimini and Weber~\cite{GRW}, appeared on the market. Currently, two sophisticated versions are intensively investigated both from the fundamental point of view as well as from the experimental point of view. These are the QMUPL model (Quantum Mechanics with Universal Position Localization)~\cite{Diosi1989} and the CSL model (Continuous Spontaneous Localization)~\cite{Pearle1989, Ghirardi1990} in its mass-proportional version~\cite{GhirardiGrassiBenatti1995}. Both models introduce the spontaneous collapse by a modification of the standard Schr\"{o}dinger equation by adding specific nonlinear and stochastic terms and introducing two new -- taking collapse models seriously -- natural constants: a collapse rate and a coherence length of the localization.

Flavor physics is a rich field within physics with many unique features, and new facilities in the near future will tackle very precisely this regime of energy. This contribution investigates how these two collapse models change the flavor changing dynamics of neutral mesons. In particular, we find that the specific link between the position space and the flavor space provides constraints onto the collapse models in general. Moreover, we find also that the noise field underlying any collapse models can be a source of the two different decay constants of mesons. In detail, the non-linear interaction of the quantum system with the noise field implies the dependence on the absolute masses of the lifetime eigenstates, the diagonal states of the Hamiltonian, such that an effectively non-unitary evolution occurs. The mathematical background are the correlation functions of the stochastic processes in the expansion. The physical background is that a time symmetric approach between the two dual spaces (bra/ket) is not necessarily required.

The QMUPL model has been investigated for the spontaneous radiation emission from a non-relativistic free charged particle~\cite{BassiDuerr2009, BassiDonadi2014} and put to an intensive experimental test by X-rays~\cite{CurceanuHiesmayrPiscicchia2015, Curceanu2015, Curceanu3}. For the mass-proportional CSL model experiments with optomechanical cavities have been proposed to detect possible changes in the spectrum of light which drives a mechanical oscillator~\cite{BahramiPaternostro2014, NimmrichterHornberger2014}. In another approach a possible increase of equilibrium temperature of a mechanical oscillator produced by the spontaneous collapse was revealed~\cite{Diosi2015}. For neutral mesons (K-, B-, D-meson) and neutrinos up to first order in time the effect of the mass-proportional CSL model was derived and compared to decoherence models~\cite{Donadi2013, Bahrami2013} by comparing to experimental data~\cite{experiments1,experiments2,experiments3,experiments4,experiments5,experiments6}. Recently, upper bounds on collapse models have been derived for cold-atom experiments~\cite{experiments7} and the authors of Ref.~\cite{darkenergy} have shown that reduction models can lead to nontrivial contribution to an effective cosmological constant.

The paper is organized as follows. We start by an introduction into collapse models and the flavor phenomenology and set the stage in Section~\ref{sec:CollapseDynamics}. Next we show how for both collapse models the transition probabilities are computed up to second order in time in Section~\ref{sec:Derivation}. These computations are lengthy and involved, therefore, the details are given in Appendix~\ref{sec:QMUPL_Computations} for the QMUPL model and in Appendix~\ref{sec:CSL_Computations} for the CSL model, respectively. The correlation functions and their dependence on the physics of the noise field are derived in Appendix~\ref{sec:TimeIntegrals}. In Section~\ref{sec:Results} we present the results, the probabilities for the lifetime states and the flavor oscillating probabilities. These are the quantities that are well investigated experimentally. We analyze then different possibilities, one allowing us an independent prediction of the effective collapse rate for the different types of neutral mesons. We proceed by giving a physical meaning to the dependence on the correlation functions of the Wiener process and finalize by developing a decoherence model that leads to the same probabilities as the CSL model, however, relies on strictly different physics. Last but not least we provide a summary and an outlook in Section~\ref{sec:Summary}.

\section{Collapse dynamics and neutral meson phenomenology}\label{sec:CollapseDynamics}
In the QMUPL and CSL models the collapse is a continuous process which can be described by the following non-linear stochastic modification of a Schr\"{o}dinger equation for a given Hamiltonian $\hat{H}$~\cite{BassiDuerrHinrichs2011}
\begin{eqnarray}\label{StateVectorEquation}
d|\phi_{t}\rangle&=&\Bigl[-i\hat{H}\, dt+\sqrt{\lambda}\sum\limits_{i=1}^N(\hat{A}_i - \langle \hat{A}_i \rangle_t) dW_{i, t} \nonumber\\
&& -\frac{\lambda}{2} \sum\limits_{i=1}^N(\hat{A}_i- \langle \hat{A}_i \rangle_t)^2dt\Bigr]|\phi_{t}\rangle,
\end{eqnarray}
with $\hbar=1$ and $\langle \hat{A}_i \rangle_t := \langle \phi_t | \hat{A}_i |\phi_{t}\rangle$ being the standard quantum mechanical expectation value. Here $\hat{A}_i$ are a set of $N$ self-adjoint commuting operators introducing the collapse in a certain basis choice (position basis in most cases), $W_{i,t}$ present a set of $N$ independent standard Wiener processes, one for each collapse operator $\hat{A}_i$, and $\lambda\geq 0$ quantifies the strength of the collapse processes. The main difference of the two collapse models under investigation is the choice of the localization operators $\hat{A}_i$.

A very useful mathematical property of equation (\ref{StateVectorEquation}) is that its physical predictions concerning the outcomes of measurements are, in terms of statistic expectations or probabilities, invariant under a phase change in the noise~\cite{AdlerBassi2007, SandroPhD}. In particular, choosing the phase to be $i$ one can rewrite the above equation by
\begin{eqnarray}\label{phaseC}
d|\phi_{t}\rangle&=&\Bigl[-i\hat{H} dt+i\;\sqrt{\lambda}\sum\limits_{i=1}^N\hat{A}_i\, dW_{i, t} -\frac{\lambda}{2} \sum\limits_{i=1}^N\hat{A}_i^2dt\Bigr]|\phi_{t}\rangle\;.\nonumber\\
\end{eqnarray}
Within collapse models statistics of outcomes of any experiment have to be expressed as averages $\mathbb{E}[\langle \phi_t|\hat{\mathbb{O}}|\phi_t\rangle]=\operatorname{Tr}[\hat{\mathbb{O}}\;\mathbb{E}[|\phi_t\rangle\langle\phi_t|]]=\operatorname{Tr}[\hat{\mathbb{O}}\;\hat{\tilde{\rho}}(t)]$, where $\hat{\mathbb{O}}$ is a self-adjoint operator.

The white noise $dW$ represents the change in time $t$ of the Wiener process $W_t$ (with the definition $W_{t=0}=1$). The term white (uncolored) refers to independent and identically distributed growths of $dW$, with a zero expectation value and a standard deviation proportional to $\sqrt{dt}$. Since the temporal derivative of a Wiener process does not exist (only in the sense of distributions) the integration of the differential equation depends on the choice of sampling point in the interval $[t,t+dt]$. The It\^{o} formalism chooses $t$ (left-hand endpoint of each time subinterval), whereas the Stratonovich formalism chooses $t+dt/2$. The advantage of the Stratonovich formalism is that the differential and integration procedures are those familiar from ordinary calculus. Therefore, we will stick to this formalism. Then equation~(\ref{phaseC}) becomes a Schr\"{o}dinger-like equation (linear) with a random Hamiltonian
\begin{equation}\label{Schroedingerlike}
\nonumber i \frac{d}{dt} |\phi_{t}\rangle = \Bigl[ \hat{H} - \sqrt{\lambda} \sum\limits_{i=1}^N \hat{A}_i w_{i, t} \Bigr] |\phi_{t}\rangle := \Bigl[ \hat{H} + \hat{N}(t) \Bigr] |\phi_{t}\rangle,
\end{equation}
where $w_{i,t}:=\frac{d}{dt} W_{i,t}$.

Flavor oscillating systems such as the neutral meson systems are described by two-state phenomenological Hamiltonians. Mesons are massive systems that decay with two (different) decay constants, consequently, giving rise to a non-unitary time evolution. To account for the decay one adds a non-Hermitian part $\hat{\Gamma}$ to the Hamiltonian. Since it is clear how to add the decay constants to the final formulae, we will later omit them in the computation of the effect of a spontaneous collapse.

In Ref.~\cite{BGHOpenSystem} the authors succeeded to show that effect of the non-hermitian part of the Hamiltonian (decay) can be understood if the system is considered to be an open quantum system. Then the Schr\"odinger equation is turned to a Gorini-Kossakowski-Lindblad-Sudarshan master equation~\cite{GKLS}, where a Lindblad operator implies the transition from the surviving part to the decaying part of the system under investigation. Consequently, the decay property can be incorporated via a Lindblad operator into the quantum system and can be physically understood as an interaction with a (virtual) environment (in quantum field theory it would refer to the QCD vacuum). This in turn shows that the total time evolution is a completely positive map. We will give the details in Section~\ref{resultsGKLS}.

A neutral meson $M^0$ consists of quark-antiquark pair, and both the particle state $|M^0\rangle$ and the antiparticle state $|\bar M^0\rangle$ can decay into the same final states. Therefore, neutral mesons have to be considered as a two-state system. The dynamics of a $M^0-\bar{M}^0$ oscillating system is covered by an effective Schr\"{o}dinger equation
\begin{align}
 & \frac{d}{dt} |\psi_t\rangle = - i\, \hat{H}_{eff} |\psi_t\rangle, \\
 & |\psi_t\rangle = a(t) |M^0\rangle + b(t) |\bar{M}^0\rangle,
\end{align}
where the phenomenological (effective) Hamiltonian $\hat{H}_{eff} = \hat{M} + \frac{i}{2} \hat{\Gamma}$ is non-hermitian, $\hat{M}=\hat{M}^\dagger$ is the mass operator which describes the unitary part of the dynamics of a neutral meson, and $\hat{\Gamma}=\hat{\Gamma}^\dagger$ covers the decay (non-unitary part). Diagonalizing the phenomenological Hamiltonian leads to two different mass eigenstates ($c=1$)
\begin{align}
 & \hat{H} |M_i\rangle = \Bigl(m_i + \frac{i}{2} \Gamma_i \Bigr) |M_i\rangle,
\end{align}
These two states $|M_L\rangle$ and $|M_H\rangle$ have distinct masses and without loss of generality $m_L$ denotes the lower one ($L$\dots light, $H$\dots heavy). For all types of neutral mesons the decay rates $\Gamma_L, \Gamma_H$ are approximately equal, except for K-mesons whose decay rates differ by a huge factor $600$. For the sake of simplicity we assume that mass eigenstates are orthogonal, $\langle M_H|M_L\rangle= 0$, herewith we neglect a small violation of the charge-conjugation--parity ($\mathbbm{CP}$) symmetry. The relation between the flavor eigenstates and mass eigenstates is then given by (introducing without loss of generality a particular phase convention)
\begin{subequations}
 \begin{eqnarray}
  & |M^0\rangle = \frac{1}{\sqrt{2}} \Bigl( |M_H\rangle + |M_L\rangle \Bigr), \\
  & |\bar{M}^0\rangle = \frac{1}{\sqrt{2}} \Bigl( |M_H\rangle - |M_L\rangle \Bigr),
 \end{eqnarray}
\end{subequations}
Let us here also remark that temporal part of the evolution of mesons is not normalized in time (due to the non-Hermitian part of the effective Hamiltonian), i.e. (for clarity we add here the natural constants which we else set to one)
\begin{eqnarray}
\psi(t)&=& e^{-\frac{i}{\hbar} m c^2\cdot t}\cdot e^{-\frac{\Gamma}{2}\cdot t}\quad\longrightarrow \int_0^\infty |\psi(t)|^2 dt\;=\;\frac{1}{\Gamma}\;.\nonumber\\
\end{eqnarray}
Obviously, a normalization of the temporal part by $\sqrt{\Gamma}$ would give a similar expression as the Born rule for the spatial part ($\int_{-\infty}^{\infty} |\psi(\vec{x})|^2 d^3x=1$) and allow for a definition of a time operator~\cite{D1,D2,D3,D4,D5,V1}. However, recently, it was shown that this formal normalization leads to contradiction with experimental data if the small violation of the charge-conjugation--parity ($\mathbbm{CP}$) symmetry is taken into account~\cite{TimeOperatorCP}. This expresses the strikingly different roles of time and space in the quantum theory and the importance of discrete symmetries.

Let us now move to the two dynamical reduction models, the QMUPL model for one particle ($N=1$) and the mass-proportional CSL model, and apply them to meson systems. In the QMUPL model the operators $\hat{A}_i$ introduce the collapse and are chosen to be the position operators $\hat{q}_i$. In order to describe the collapse dynamics in the case of neutral mesons we extend the collapse operators $\hat{A}_i$ by a flavor part
\begin{subequations}
\begin{widetext}
 \begin{eqnarray}\label{CollapseOperatorQMUPL}
 & \hat{\mathbf{A}}_{QMUPL} = \hat{\mathbf{q}} \otimes \Bigl[ \frac{m_H}{m_0} | M_H \rangle \langle M_H | + \frac{m_L}{m_0} | M_L \rangle \langle M_L | \Bigr],
 \end{eqnarray}
and, consequently, the $\hat{N}(t)$ operator of the Schr\"{o}dinger-like equation~(\ref{Schroedingerlike}) becomes
\begin{eqnarray}
 & \hat{N}_{QMUPL}(t) = -\sqrt{\lambda} \Bigl( \mathbf{w}(t) \cdot \hat{\mathbf{q}} \Bigr) \otimes \Bigl[ \frac{m_H}{m_0} | M_H \rangle \langle M_H | + \frac{m_L}{m_0} | M_L \rangle \langle M_L | \Bigr], \label{PerturbationQMUPL}
 \end{eqnarray}
\end{widetext}
\end{subequations}
where $m_0$ is a reference mass which is taken usually to be the nucleon mass. We consider $\mathbf{w}(t):=\frac{d\mathbf{W}(t)}{dt}$ as a white (uncolored) noise field, where $\mathbf{W}(t) = \{ W_1 (t), ..., W_d (t) \}$ and the corresponding correlation function is $\mathbb{E}[\mathbf{w}(t) \cdot \mathbf{w}(s)] = \delta(t-s)$. Here $d$ denotes the dimension, i.e. $d=1,2,3$ in general.

In the case of the mass-proportional CSL model the collapse operator acts in a Fock space, so we replace $\hat{A}_i$ by a continuous set of operators $\hat{A} (\mathbf{x})$, one for each point in space, i.e.
\begin{widetext}
\begin{subequations}
 \begin{eqnarray}
 & \hat{A}_{CSL} (\mathbf{x}) = \displaystyle\int d\mathbf{y}\; g(\mathbf{y}-\mathbf{x}) \Bigl(\frac{m_H}{m_0} \hat{\psi}^{\dagger}_H (\mathbf{y})\hat{\psi}_H (\mathbf{y}) + \frac{m_L}{m_0} \hat{\psi}^{\dagger}_L (\mathbf{y})\hat{\psi}_L (\mathbf{y}) \Bigr), \label{CollapseOperatorCSL} \\
 & \hat{N}_{CSL}(t) = -\sqrt{\gamma} \displaystyle\int d\mathbf{y} \; w(\mathbf{y},t) \Bigl( \frac{m_H}{m_0} \hat{\psi}^{\dagger}_H (\mathbf{y}) \hat{\psi}_H (\mathbf{y}) + \frac{m_L}{m_0} \hat{\psi}^{\dagger}_L (\mathbf{y}) \hat{\psi}_L (\mathbf{y}) \Bigr), \label{PerturbationCSL}
 \end{eqnarray}
\end{subequations}
\end{widetext}
where $\hat{\psi}^{\dagger}_j(\mathbf{y})$ and $\hat{\psi}_j(\mathbf{y})$ are creation and annihilation operators of a particle of type $j=H,L$ in a point $\mathbf{y}$. The smearing function $g(\mathbf{y}-\mathbf{x})$ is usually taken to be of a Gaussian type
\begin{equation}
g(\mathbf{y}-\mathbf{x}) = \frac{1}{(\sqrt{2\pi}r_C)^d}\; e^{-(\mathbf{y}-\mathbf{x})^2/2r_C^2},
\end{equation}
where $r_C$ is a spatial correlation length and represents one of the two phenomenological constants $\gamma,r_c$ of the mass-proportional CSL model. The correlation functions of the mass-proportional CSL noise are given by \begin{eqnarray}\mathbb{E}[w (\mathbf{x},t) w (\mathbf{y},s)] = F(\mathbf{x}-\mathbf{y})\delta(t-s)\;,\end{eqnarray} where $F(\mathbf{x}) = \frac{1}{(\sqrt{4\pi}r_C)^d} e^{-\mathbf{x}^2/4r_C^2}$. Note that we have substituted the rate $\lambda$ by $\gamma$ which has now the units $[m^d/s]$. A characteristic of the CSL model is that all observable results will be proportional to the ratio $\gamma/r_C^d$ being a rate or by including all units the strength of the interaction.

\section{Derivation of the transitions probabilities}\label{sec:Derivation}

Accelerator facilities have intensively studied and will study the transition probabilities from mass eigenstates to mass eigenstates, $P_{M_\mu \rightarrow M_\nu}(t)$, and from flavor eigenstates to flavor eigenstates, $P_{M^0 \rightarrow M^0/\bar{M}^0}(t)$.

For the QMUPL model we need to define the initial spatial state. We will assume a wave packet in position picture with a width $\sqrt{\alpha}$ in $d$-dimensional space ($d=1,2,3$) and with a momentum $\mathbf{p}_i$. Further we assume that the final state is a momentum eigenstate (most common scenario in typical accelerator facilities). The probabilities of interest are
\begin{subequations}
\begin{align}
& \nonumber P_{M_\mu \rightarrow M_\nu}(\alpha ; t) = \sum_{\mathbf{p}_f} \mathbb{E} \left| \langle M_\nu, \mathbf{p}_f | M_\mu (t) , \mathbf{p}_i, \alpha \rangle \right|^2, \\
& \nonumber P_{M^0 \rightarrow M^0/\bar{M}^0}(\alpha ; t) = \sum_{\mathbf{p}_f} \mathbb{E} \left| \langle M^0/\bar{M}^0, \mathbf{p}_f | M^0 (t), \mathbf{p}_i, \alpha \rangle \right|^2,
\end{align}
\end{subequations}
where $\mathbb{E}$ denotes the noise average and $\mu,\nu = L,H$. For the QMUPL model we start with the $1$-dimensional case and then generalize the results to the $d$-dimensional case. In the case of the mass-proportional CSL model we start directly with the $d$-dimensional case.

To obtain the probabilities of interest we need to compute first the transition amplitudes for all mass eigenstates. For that we move to the interaction picture and treat the noise term $\hat{N}(t)$ as a perturbation
\begin{align}
& T_{\mu\nu}(\mathbf{p}_f, \mathbf{p}_i, \alpha ; t) := \langle M_{\nu}, \mathbf{p}_f | M_{\mu} (t), \mathbf{p}_i, \alpha \rangle \\
& \nonumber = e^{-i m_{\mu} t}\; \langle M_{\nu}, \mathbf{p}_f | \hat{U}_I(t) | M_{\mu}, \mathbf{p}_i, \alpha \rangle,
\end{align}
where the evolution operator $\hat{U}_I(t)$ is the corresponding one in the interaction picture. The evolution operator is then expanded into a Dyson series up to fourth perturbative order
\begin{align}\label{DysonComponents}
& T_{\mu \nu}(\mathbf{p}_f, \mathbf{p}_i, \alpha ; t) \simeq e^{-i m_{\mu} t} \Bigl( T^{(0)}_{\mu\nu}(\mathbf{p}_f, \mathbf{p}_i, \alpha ; t)  \\
& \nonumber + T^{(1)}_{\mu\nu}(\mathbf{p}_f, \mathbf{p}_i, \alpha ; t) + T^{(2)}_{\mu\nu}(\mathbf{p}_f, \mathbf{p}_i, \alpha ; t) \\
& \nonumber + T^{(3)}_{\mu\nu}(\mathbf{p}_f, \mathbf{p}_i, \alpha ; t) + T^{(4)}_{\mu\nu}(\mathbf{p}_f, \mathbf{p}_i, \alpha ; t)\Bigr),
\end{align}
with
\begin{subequations}
\begin{align}
 & T^{(0)}_{\mu\nu}(\mathbf{p}_f, \mathbf{p}_i, \alpha ; t_0) = \langle M_{\nu}, \mathbf{p}_f | M_{\mu}, \mathbf{p}_i, \alpha \rangle, \label{QMUPL_AmpZeroOrder} \\
 & T^{(n)}_{\mu\nu}(\mathbf{p}_f, \mathbf{p}_i, \alpha ; t_0) = (-i)^n \int_0^{t_0} dt_1 ... \int_0^{t_{n-1}} dt_n \label{QMUPL_AmpFourthOrder} \\
 & \nonumber \;\;\;\;\; \cdot \langle M_{\nu}, \mathbf{p}_f | \prod\limits_{j=1}^n \Bigl( \hat{N}_I (t_j) \Bigr) | M_{\mu}, \mathbf{p}_i, \alpha \rangle \; \mbox{for} \; n=1,2,3,4,
\end{align}
\end{subequations}
where $\hat{N}_I(t)$ is the noise term in the interaction picture. Each term in~(\ref{DysonComponents}) is in detail computed in the appendix~\ref{sec:QMUPL_Computations}. The crucial derivations of the correlation functions are given in appendix~\ref{sec:TimeIntegrals}.

For the CSL model we follow a similar strategy, however, we can immediately consider the $d$-dimensional case and the method introduced in Ref.~\cite{Donadi2013, SandroPhD}. All details are summarized in appendix~\ref{sec:CSL_Computations}.

\section{Results}\label{sec:Results}

We first give the general result and analyze its features. Then we present a decoherence model within standard quantum mechanics resulting in the same probabilities as those of the CSL model.

\subsection{General results}

Putting all pieces together we obtain the desired probabilities up to second order in time and collapse parameters
\begin{widetext}
\begin{align}
& P^{QMUPL}_{M_{\mu=L/H}\rightarrow M_{\nu=L/H}} (t) = \delta_{\mu\nu}\;\Bigl( 1 -  \Lambda^{QMUPL}_\mu\cdot t + 3\cdot\frac{1}{2}(\Lambda^{QMUPL}_\mu)^2\cdot t^2 +O(t^3)\Bigr)\cdot e^{-\Gamma_\mu t}, \\
& P^{CSL}_{M_{\mu=L/H}\rightarrow M_{\nu=L/H}} (t) = \delta_{\mu\nu}\;\Bigl( 1 -  \Gamma^{CSL}_\mu\cdot  t + \frac{1}{2}(\Gamma^{CSL}_\mu)^2 \cdot t^2 +O(t^3)\Bigr)\cdot e^{-\Gamma_\mu t},
\end{align}
\end{widetext}
with \begin{eqnarray}\Lambda^{QMUPL}_\mu&=&\frac{\alpha\lambda}{2}\cdot \frac{m_{\mu}^2}{m_0^2}\cdot  \Bigl( 1 - 2\theta(0)\Bigr)\;,\nonumber\\
\Gamma^{CSL}_\mu&=& \frac{\gamma}{(\sqrt{4\pi}r_C)^d}\cdot \frac{m_{\mu}^2}{m_0^2}\cdot \Bigl(1 - 2\theta(0)\Bigr)\;,\end{eqnarray} where $\theta(0)$ is the Heaviside function at zero. The result of the QMUPL model agrees with those in Ref.~\cite{SimonovHiesmayr2016}. Let us mention here a couple of comments. Firstly, the mass eigenstates do not oscillate as it is the case in the standard approach. Secondly, we find that the effect of the collapse in position space leads to a term that is proportional to the mass squared per unit mass squared. These masses never appear in standard quantum theory. Moreover, it gives an ``inverted'' ordering, namely the decay rate that is bigger than the other one is connected to the heavier mass. This in turns means that the eigenstate of the heavier mass decays earlier. We reconsider this point in section~\ref{fulldynamics}.
Thirdly, the result of the QMUPL model is independent of the number of dimensions $d$ (see appendix~\ref{sec:CSL_Computations}). Fourthly, there is an additional factor $3$ (independent of the number of dimensions $d$) in second order of time. This is a bit surprising. Having a closer look into the computations summarized in appendix~\ref{sec:QMUPL_Computations} we observe that the factor $3$ is a product of choosing Gaussian wave functions and their integration over all final momenta. As a consequence the effect of the collapse on the meson time evolution cannot be assumed to be an exponential effect in general. In strong contrast to the CSL model, where we can expect that the dynamics of a mass eigenstate propagating in free space is exponential
\begin{eqnarray}\label{finalmassdynamics}
P^{CSL}_{M_{\mu=L/H}\rightarrow M_{\nu=L/H}} (t)&=& \delta_{\mu\nu}\; e^{-(\Gamma^{CSL}_\mu+\Gamma_\mu) t}\;.
\end{eqnarray}
Last but not least the choice of $\theta(0)\in[0,1]$ coming from the correlation functions of the Wiener processes leads to positive ($\theta(0)\in[0,\frac{1}{2 }\}$), zero ($\theta(0)=\frac{1}{2}$) or negative ($\theta(0)\in[\frac{1}{2},1\}$) values of $\Gamma^{CSL}$. Before we proceed in discussing which value $\theta(0)$ should be taken, let us present the results for the transition of the flavor eigenstates (flavor oscillation)
\begin{widetext}
\begin{eqnarray}
P^{QMUPL}_{M^0 \rightarrow M^0/\bar{M}^0} &=& \frac{1}{4} \Biggl\{ \sum_{i=H,L} e^{-\Gamma_i t}\; \Bigl( 1 -  \Lambda^{QMUPL}_i\cdot t + 3\cdot\frac{1}{2}(\Lambda^{QMUPL}_i)^2\cdot t^2 +O(t^3)\Bigr)\nonumber\\
 && \pm 2\cos(\Delta m t)\;e^{-\frac{\Gamma_H+\Gamma_L}{2} t} \cdot \left(1-\frac{\alpha \lambda}{2} \left[\frac{\Delta m^2}{m_0^2} (1-\theta(0)) + \frac{m_H m_L}{m_0^2} (1-2 \theta(0))\right]\cdot t\right.\nonumber\\
 &&\left.\qquad + \; 3\cdot\frac{1}{2}\left(\frac{\alpha \lambda}{2} \left[\frac{\Delta m^2}{m_0^2} (1-\theta(0))+ \frac{m_H m_L}{m_0^2} (1-2 \theta(0))\right]\right)^2\cdot t^2+O(t^3)\right)\Biggr\},\nonumber\\
P^{CSL}_{M^0 \rightarrow M^0/\bar{M}^0} &=& \frac{1}{4} \Biggl\{ \sum_{i=H,L} e^{-\Gamma_i t}\; \Bigl( 1 -  \Gamma^{CSL}_i\cdot t + \frac{1}{2}(\Gamma^{CSL}_i)^2\cdot t^2 +O(t^3)\Bigr)\nonumber\\
 && \pm 2\cos(\Delta m t)\;e^{-\frac{\Gamma_H+\Gamma_L}{2} t}\cdot \left(1-\frac{\gamma}{(\sqrt{4\pi}r_C)^d} \left[\frac{\Delta m^2}{m_0^2} (1-\theta(0))+\frac{m_H m_L}{m_0^2} (1-2 \theta(0))\right]\cdot t\right.\nonumber\\
 &&\left.\qquad +\frac{1}{2}\left(\frac{\gamma}{(\sqrt{4\pi}r_C)^d} \left[\frac{\Delta m^2}{m_0^2} (1-\theta(0))+\frac{m_H m_L}{m_0^2} (1-2 \theta(0))\right]\right)^2\cdot t^2+O(t^3)\right)\Biggr\},
\end{eqnarray}
where $\Delta m = m_H - m_L$. Again, for the CSL model we assume that the higher orders in time lead to an exponential behavior
\begin{eqnarray}
\lefteqn{P^{CSL}_{M^0 \rightarrow M^0/\bar{M}^0}(t)\;=}\\
 &&\frac{1}{4} \Biggl\{ e^{-(\Gamma_H+\Gamma^{CSL}_H)t}+e^{-(\Gamma_L+\Gamma^{CSL}_L)t}\pm 2\cos(\Delta m t)\;e^{-\frac{\Gamma_H+\Gamma_L}{2} t}\cdot e^{-\frac{\Gamma_H^{CSL}+\Gamma_L^{CSL}}{2} t}\cdot e^{- \frac{\gamma}{ (\sqrt{4\pi} r_C)^d}\frac{(\Delta m)^2}{2 m_0^2} t} \Biggr\}\;.
\end{eqnarray}

This we can rewrite in the following form
\begin{eqnarray}\label{finalresultprob}
P^{CSL}_{M^0 \rightarrow M^0/\bar{M}^0}(t) &=& \frac{e^{-(\Gamma_L+\Gamma^{CSL}_L)t}+e^{-(\Gamma_H+\Gamma^{CSL}_H)t}}{4}\cdot 
\Biggl\{ 1\pm \frac{\cos(\Delta m t)}{\cosh(\frac{(\Gamma_L+\Gamma^{CSL}_L)-(\Gamma_H+\Gamma^{CSL}_H)}{2}\cdot t)}\cdot e^{- \frac{\gamma}{ (\sqrt{4\pi} r_C)^d}\frac{(\Delta m)^2}{2\,m_0^2} t} \Biggr\}\;.
\end{eqnarray}
\end{widetext}
This is an interesting result since it disentangles two effects of the collapse model. A damping of the interference term proportional to the mass difference squared $(\Delta m)^2$ is independent of the choice of the Heaviside function $\theta(0)$ and additional energy terms $\Gamma^{CSL}_i$ proportional to the absolute masses depend on the Heaviside function. These additional energy terms play the same role as the decay constants (added by hands) in standard quantum theory. In the next step we investigate whether the collapse dynamics leading to the above result can explain the full dynamics of the neutral meson systems without defining decay constants (by hands) due to Wigner-Weisskopf approximation.

\subsection{Impact and observability of the CSL model prediction}\label{fulldynamics}

At accelerator facilities the following asymmetry term $A(t)$ is experimentally well investigated
\begin{eqnarray}\label{asymmetry}
A(t)&=&\frac{P^{CSL}_{M^0 \rightarrow M^0}(t)-P^{CSL}_{M^0 \rightarrow \bar{M}^0}(t)}{P^{CSL}_{M^0 \rightarrow M^0}(t)+P^{CSL}_{M^0 \rightarrow \bar{M}^0}(t)}\\
&=&\frac{\cos(\Delta m t)}{\cosh(\frac{(\Gamma_L+\Gamma^{CSL}_L)-(\Gamma_H+\Gamma^{CSL}_H)}{2}\cdot t)}\cdot e^{- \frac{\gamma}{ (\sqrt{4\pi} r_C)^d}\frac{(\Delta m)^2}{2\,m_0^2} t}\;.\nonumber
\end{eqnarray}
From that we observe that the damping term proportional to $\frac{\gamma}{ (\sqrt{4\pi} r_C)^d}\frac{(\Delta m)^2}{2\,m_0^2}$ is in principle measurable. The standard proposed value for the mass-proportional CSL model is $\lambda_{CSL}:=\frac{\gamma}{(\sqrt{4\pi} r_C)^d}\approx 10^{-(8\pm2)}s^{-1}$ (Adler~\cite{Adler}) or $\approx 10^{-16}s^{-1}$ (GRW~\cite{GRW}). Here the coherence length is assumed to be of the order $10^{-5}cm$ and $d=3$ and from that the collapse strength $\gamma$ can be deduced. For more details on the allowed parameter space for $r_C$ and $\gamma$ consider, e.g., Ref.~\cite{FeldmannTumulka}. Let us also note that the best experimental upper bound is currently obtained by $X$-rays~\cite{CurceanuHiesmayrPiscicchia2015} being five orders away from the proposed value of Adler, i.e. $10^{-12} s^{-1}$.

Plugging in these two values (Adler/GRW) and the measured mass differences we find damping rates of the order $10^{-38}s^{-1}/10^{-46}s^{-1}$ for $K$-mesons, $10^{-30}s^{-1}/10^{-38}s^{-1}$ for $B_d$-mesons, $10^{-30}s^{-1}/10^{-38}s^{-1}$ for $B_s$ mesons and $10^{-34}s^{-1}/10^{-42}s^{-1}$ for $D$ mesons (see also Ref.~\cite{Bahrami2013}). The choice of the reference mass $m_0$ being either the neutron mass or the rest mass of the respective neutral meson does not affect the values considerably.  This is not directly observable since it corresponds to a lifetime much greater than the decay rates of the respective neutral meson. Consequently, the effect of the spontaneous collapse on the interference can be safely neglected.

The idea behind the choice of $m_0$, being generally a free parameter of the CSL model, is that for ordinary matter the mass ratio corresponds to an average number of constituents of the composite object~\cite{PearleSquires}; the bigger the object, the stronger the effect of spontaneous localization. The choice in the meson system stems from our assumption that if collapse models are relevant in Nature then they have to hold for all physical systems. For the meson system this mass ratio $\frac{m_\mu}{m_0}$ decreases if $m_0$ is of the order of a nucleon or the rest mass of the mesons system, i.e. has the opposite behavior. Thus it may seem more reasonable to have for particles lighter than those that make up the ordinary matter the inverted ratio. If we do so then the damping factor of the interference term becomes
$\frac{1}{2} \lambda_{CSL} \frac{\Delta m^2 m_0^2}{m_H^2 m_L^2}$, which is only computable if we know the absolute masses.


The second modification due to the mass-proportional CSL model compared to the standard approach is for the decay rates, i.e. $\Gamma_\mu+\Gamma^{CSL}_\mu$. Here $\Gamma_\mu$ are the standard decay rates introduced to the system by the Wigner-Weisskopf approximation. The collapse contribution is connected to the absolute mass (playing no role in the standard approach) and the value of the Heaviside function at zero, i.e. $\Gamma^{CSL}_\mu=\lambda_{CSL}\cdot \frac{m_{\mu}^2}{m_0^2}\cdot \Bigl(1 - 2\theta(0)\Bigr)$ or in the inverted scenario $\Gamma^{CSL}_\mu=\lambda_{CSL}\cdot \frac{m_{0}^2}{m_\mu^2}\cdot \Bigl(1 - 2\theta(0)\Bigr)$.

Taking this one step further is to ask whether collapse models could solely be responsible for the decaying part of the neutral mesons, i.e. the dynamics of the spontaneous localization induces the decay of the mass eigenstates. For that we set $\Gamma_\mu^{exp}\equiv\Gamma^{CSL}_\mu$. Certainly $\Gamma^{CSL}$ needs to be positive, i.e $\theta(0)\in[0,\frac{1}{2}\}$, to obey equations (\ref{finalmassdynamics}). Then we obtain
\begin{widetext}
\begin{eqnarray}\label{solintr}
\frac{\Gamma^{CSL}_L-\Gamma^{CSL}_H}{\Gamma^{CSL}_L+\Gamma^{CSL}_H}&\stackrel{\tiny{\theta(0)\not=\frac{1}{2}}}{=}&\pm\frac{m_L^2-m_H^2}{m_L^2+m_H^2}=\left\lbrace\begin{array}{l}
\textrm{K-mesons: }0.996506\left\lbrace\begin{array}{l} +1.2760\cdot 10^{-5}\\
-1.2760\cdot 10^{-5}\end{array}\right.\\
\textrm{D-mesons: } 0.00645 \left\lbrace\begin{array}{l} +0.0007\\
-0.0009\end{array}\right.\\
\textrm{$B_d$-mesons: } 0.0005\left\lbrace\begin{array}{l} +0.0050\\
-0.0050\end{array}\right.\\
\textrm{$B_s$-mesons: } 0.06912\left\lbrace\begin{array}{l} +7.7058\cdot 10^{-4} \\
-7.7058\cdot 10^{-4} \end{array}\right.\\
\end{array}\right.
\end{eqnarray}
The experimental values for the experimentally measured decay constants (right--hand side of the above equation) are taken from the particle data book~\cite{ParticleDataBook}. The method how to deduce from the experimental values measured the decay rates is described in appendix~\ref{numerics} since it differs slightly for each meson. The minus sign holds for the inverted scenario. Together with the experimentally obtained value of $\Delta m:=m_H-m_L$, this allows to compute the absolute values of the masses $m_{H/L}$ via
\begin{eqnarray}
\frac{\Gamma^{CSL}_L-\Gamma^{CSL}_H}{\Gamma^{CSL}_L+\Gamma^{CSL}_H}\;=\;\pm\frac{m_L^2-(m_L+\Delta m)^2}{m_L^2+(m_L+\Delta m)^2}\;=\;\;\pm \frac{-2 m_L\Delta m-(\Delta m)^2}{2 m_L^2+2 m_L\Delta m+(\Delta m)^2}\;=\;\pm\left(- 1+\frac{m_L^2}{m_L^2+m_L\Delta m+\frac{1}{2}(\Delta m)^2}\right)\;.\nonumber\\
\end{eqnarray}

\end{widetext}
In the case we have $m_H> m_L$ ($\Delta m>0$) we observe that the right--hand side becomes negative (if we do not reverse the mass ratio). Thus the two involved masses cannot be both positive. This is because the collapse models relate the decay rates with the corresponding masses directly proportionally: the heavier the mass the bigger the decay rate, the smaller the lifetime. This is physically intuitive from the collapse model perspective since heavier masses should be affected stronger by the spontaneous factorization. The counter-intuitive effect for applying that to neutral mesons decay is that the more massive state should decay faster. In literature there can be found experiments~\cite{sign1,sign2} for K-mesons assigned to measure the sign of $\Delta m$ and, herewith, if the heavier mass connects also to the lower decay rate (longer lifetime) and vice versa.  The results are a positive sign of $\Delta m$, i.e. the heavier mass decays slower. Note that not for all mesons the sign has been determined.  In summary, for positive mass differences $\Delta m>0$ we cannot find positive masses.

In the reversed scenario positive values for the absolute masses are obtained and listed in table~\ref{table}. The obtained values for the absolute masses are in the regime of the weak interaction due to our identification with the decay rates. They are functions of the two decay rates and the mass difference. Let us remind the reader that the rest mass of mesons is by many units higher since here the strong interaction rules. Note that the numerical values are very sensitive to the errors and the method to determine the decay constants which are very different to the specific mesons and the experiments considered. We stick here to the values published by the particle data group in their summary and review papers~\cite{ParticleDataBook}.

\begin{widetext}
\begin{center}
\begin{table}
  \begin{tabular}{ ||c | c | c | c | c | c || }
    \hline
    & $\Gamma_L^{\textrm{exp}}$ [$s^{-1}$]& $\Gamma_H^{\textrm{exp}}$  [$s^{-1}$]& $\Delta m^{\textrm{exp}}$  [$\hbar s^{-1}$]& $m_L$ [$\hbar s^{-1}$]& $m_H$ [$\hbar s^{-1}$]\\ \hline
    $K$-mesons & $1.117 \cdot 10^{10} $ & $1.955 \cdot 10^{7} $ &$0.529 \cdot 10^{10}$& $2.311 \cdot 10^8$ & $5.524 \cdot 10^9$  \\
    $D$-mesons &  $2.454\cdot 10^{12}$ & $2.423 \cdot 10^{12}$ &$0.950 \cdot 10^{10}$&  $1.468 \cdot 10^{12}$ & $1.477 \cdot 10^{12}$  \\
    $B_d$-mesons & $6.582 \cdot 10^{11}$ & $6.576 \cdot 10^{11} $ &$0.510 \cdot 10^{12}$& $1.020 \cdot 10^{15} $ & $1.020 \cdot 10^{15} $  \\
    $B_s$-mesons & $7.072\cdot 10^{11} $ & $6.158 \cdot 10^{11} $ &$ 1.776 \cdot 10^{13}$& $2.477 \cdot 10^{14} $ & $2.655 \cdot 10^{14} $ \\
    \hline
  \end{tabular}\caption{Experimental values of the decay rates, the mass difference and the computed values of the absolute masses for the neutral mesons system.}\label{table}
  \end{table}
\end{center}
\end{widetext}

\begin{figure}
\includegraphics[width=0.48\textwidth]{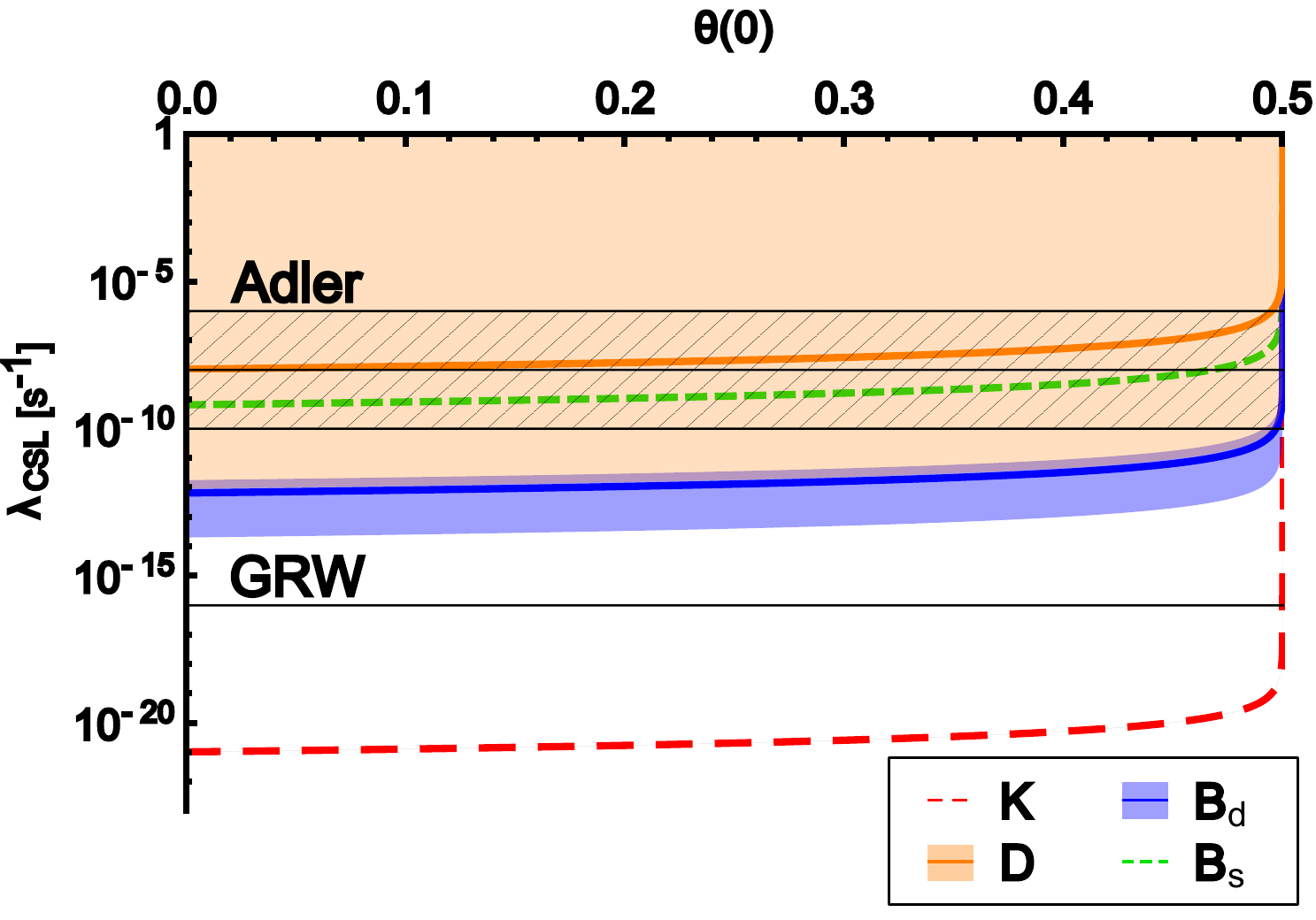}
\caption{(Color online) These plots show the values of $\theta(0)$ versus the deduced collapse rate based on the input parameters $\Gamma_H^{\textrm{exp}},\Gamma_L^{\textrm{exp}},\Delta m^{\textrm{exp}}$ for the different types of neutral mesons (including experimental errors). As a reference mass the respective rest mass of the neutral mesons is assumed.}\label{figure1}
\end{figure}

Now we can use these values of absolute masses to estimate $\lambda_{CSL}$ by
\begin{eqnarray}
\lambda_{CSL}^{\textrm{estimated}}&:=&\Gamma_\mu^{\textrm{exp}}\cdot \frac{m_\mu^2}{m_0^2}\frac{1}{(1-2\theta(0))}\\
&=&\frac{1}{\biggr(\sqrt{\Gamma_L^{-1}}-\sqrt{\Gamma_H^{-1}}\biggl)^2}\frac{(\Delta m)^2}{m_0^2}\frac{1}{(1-2\theta(0))}\nonumber\;.
\end{eqnarray}
The predicted values are plotted in Fig.~\ref{figure1} for the different meson types. Interestingly, these values correspond to the ones assumed by Adler, except for the K-meson system which is closer to the one of GRW (even weaker). Fixing the collapse rate to the one proposed by GRW requires that $\theta(0)$ converges to $\frac{1}{2}$, only in the Adler case values $\not=\frac{1}{2}$ are allowed. Taking the scenario with reversed masses seriously we have also to consider the modified contribution to the interference term, i.e.
\begin{eqnarray}
&&\frac{1}{2} \lambda_{CSL} \frac{\Delta m^2 m_0^2}{m_H^2 m_L^2}\nonumber\\
&=& \frac{1}{2} \lambda_{CSL} \frac{m_0^2}{(\Delta m)^2} \Gamma_H \Gamma_L \biggr(\sqrt{\Gamma_L^{-1}}-\sqrt{\Gamma_H^{-1}}\biggl)^4\\
&=&\frac{1}{2} \frac{\lambda_{CSL}}{\lambda_{CSL}^{\textrm{estimated}}}\frac{1}{1-2\theta(0)} \biggr(\sqrt{\Gamma_L}-\sqrt{\Gamma_H}\biggl)^2\;.\nonumber
\end{eqnarray}
This term is negligible for all types of neutral mesons  due to the tiny difference between the two decay constants assuming that the other values are of order $1$ except for the K-meson system. In this case we have a very sensitive tradeoff between obtaining the experimental values of the decay constant and the damping of the interference term. The best limit on such a possible modification of the interference term comes from the entangled K-meson system~\cite{experiments3}, however, this is not directly comparable.

In summary a full description of the decay and oscillation properties in the dynamics of neutral mesons can be obtained demanding certain properties of collapse models.

\subsection{What is the physical meaning of the choice of $\theta(0)\in[0,1]$?}

Let us remind the reader that it comes from the action of the classical noise underlying any collapse model. One assumes that the stochastic noise average of two Wiener processes is given by a delta distribution,  $\mathbb{E}[\mathbf{w}(t) \cdot \mathbf{w}(s)] = \delta(t-s)$. This in turn is the assumption of the white noise scenario, i.e. the assumption of a constant power spectral density. In our derivation we had to compute the following type of time integrals
\begin{eqnarray}
\int_0^t ds\; \delta(t-s)&=&\biggl\lbrace\begin{array}{c} \theta(t)-\theta(0)=1-\theta(0)\;\textrm{for}\; t-s\geq 0\\
\theta(0)-\theta(-t)=\theta(0)\quad\textrm{for}\; t-s\leq 0\\
\end{array}\nonumber\\
\end{eqnarray}
with $\int_{-\infty}^{\infty}\delta(t)d\,t=1$. Note that the dependence on $\theta(0)$ occurs only in case one matches amplitudes of different orders within the expansion. Assuming the independence of the time direction $\delta(t)=\delta(-t)$ leads to $\theta(0)=\frac{1}{2}$. In this case the collapse quantities $\Lambda^{QMUPL}, \Gamma^{CSL}$ become zero, respectively. No effect of the collapse field arises in the evolution of the mass eigenstates. Consequently, in this case also no dependence on absolute masses ($m_H$, $m_L$) is proposed in line with the standard quantum mechanical approach.

A value $\theta(0)\not=\frac{1}{2}$ can be interpreted as the freedom that the time evolved states in the expansions in the ``out'' (``bra'') and the ``in'' (``ket'') states do depend on the particular time ordering within the expansion. Only in this case the interaction with the classical noise field leads to contributions not solely affecting the interference term with respect to the chosen basis. Consequently, here is the point where the physics of the noise field strongly enters the discussion. In particular non-white noise fields will change the very dynamics of neutral mesons, that in turn will be testable.

\subsection{Mimicking the effect of the CSL model on the meson dynamics by a decoherence model}\label{resultsGKLS}

To better understand the physics proposed by collapse models for the meson dynamics, let us see which decoherence model within standard quantum mechanics would in principle lead to the same predictions.

Let us here also say some words about the non-hermitian Hamiltonian that is the the standard starting point in describing the meson phenomenology. Neutral meson systems violate the $\mathbbm{CP}$ symmetry for the mass matrix and have a non-vanishing lifetime difference in the width matrix. This leads to an effective Hamiltonian which is even not a normal operator with incompatible (non-commuting) masses and widths. In the Wigner-Weisskopf approach, by diagonalizing the entire Hamiltonian, the in general non-orthogonal ``\textit{stationary}'' states $M_H,M_L$ are obtained. These states have complex eigenvalues whose real (imaginary) part does not coincide with the eigenvalues of the mass (width) matrix. The mesonic systems can also be described as an open quantum mechanical system~\cite{BGHOpenSystem,Mavromatos,Smolinski}. In particular, the following Gorini-Kossakowski-Lindblad-Sudarshan master equation does the job~\cite{BGHOpenSystem}
\begin{eqnarray}
\frac{d}{dt}\hat{\rho}(t)&=& -i [\hat{\mathcal{H}},\hat{\rho}(t)]\\
&&-\frac{1}{2}\sum_{i=0}^f\left(\hat{\mathcal{L}}_i^\dagger \hat{\mathcal{L}}_i \hat{\rho}(t)+\hat{\rho}(t)\hat{\mathcal{L}}_i^\dagger \hat{\mathcal{L}}_i- 2 \hat{\mathcal{L}}_i\hat{\rho}(t) \hat{\mathcal{L}}_i^\dagger\right)\nonumber
\end{eqnarray}
where we define $\hat{\rho}$ to live on a Hilbert-Schmidt space with a direct product structure $\textsf{H}_s\oplus\textsf{H}_d$ ($s$ corresponds to the surviving part and $d$ of the decaying part of the system). In our case we need at least a four-dimensional space. The Hamiltonian $\hat{\mathcal{H}}$ and all Lindblad operators $\hat{\mathcal{L}}$ are defined to act only onto the surviving part of the system, i.e.
\begin{eqnarray}
\hat{\mathcal{H}}&=&\left(\begin{array}{cc} \hat{M}&0\\
0&0\\
\end{array}\right)\;,\quad\hat{\mathcal{L}}_{i>0}\;=\;\left(\begin{array}{cc} \hat{L}_i&0\\
0&0\\
\end{array}\right)\;,
\end{eqnarray}
whereas the zero Lindblad operator entangles the surviving part with the decaying part
\begin{eqnarray}
\hat{\mathcal{L}}_0&=&\left(\begin{array}{cc} 0&0\\
\hat{L}_0&0\\
\end{array}\right)\;.
\end{eqnarray}
Given these definition the total density matrix
\begin{eqnarray} \hat{\rho}(t)&=&\left(\begin{array}{cc} \hat{\rho}_{ss}(t)&\hat{\rho}_{sd}(t)\\
\hat{\rho}_{sd}^\dagger(t)&\hat{\rho}_{dd}(t)\\
\end{array}\right)
\end{eqnarray}
is normalized for all times. The differential equation decouples for the parts of the system. Hence, the solution of the survive-decay part $\hat{\rho}_{sd}(t)$ has no physical significance and the time dependence of the decay-decay contribution $\hat{\rho}_{dd}(t)$ depends solely on $\hat{\rho}_{ss}(t)$, i.e.
\begin{eqnarray}
\hat{\rho}_{dd}(t)&=& \hat{L}_0\;\int_0^{t} \hat{\rho}_{ss}(t') dt'\; \hat{L}_0^\dagger\;.
\end{eqnarray}

For clarity, let us rewrite the relevant differential equation explicitly (for hermitian Lindblad generators)
\begin{eqnarray}
\hat{\rho}_{ss}(t)&=& -i\,[\hat{H},\hat{\rho}_{ss}(t)]-\frac{1}{2}\{\hat{L}_0,\hat{\rho}_{ss}(t)\}\nonumber\\
&&-\frac{1}{2}\sum_{i>0} \left\lbrace\{\hat{L}_i,\hat{\rho}_{ss}(t)\} - 2 \hat{L}_i\hat{\rho}_{ss}(t) \hat{L}_i\right\rbrace
\end{eqnarray}
is given in the mass eigenstate basis by $\hat{L}_0=\textrm{diag}\{\sqrt{\Gamma_L+\Gamma_L^{CSL}},\sqrt{\Gamma_H+\Gamma_H^{CSL}}\}$. Choosing $\hat{L}_1=\sqrt{\frac{\gamma}{(\sqrt{4\pi} r_C)^d}}(\frac{m_L}{m_0}|M_L\rangle\langle M_L| + \frac{m_H}{m_0}|M_H\rangle\langle M_H|)$ formally leads to the same probabilities, see Eq.~(\ref{finalresultprob}).

This has the following physical intuitive picture behind it: the state vector undergoes a random unitary transformation in the time $dt$
\begin{eqnarray}
\hat{U}(\phi) |\psi(t)\rangle &=& e^{-i \phi \hat{G}}\;|\psi(t)\rangle\nonumber\\
&=&(\mathbbm{1}-i \phi \hat{G}-\frac{1}{2} \phi^2 \hat{G}^2+\dots)|\psi(t)\rangle
\end{eqnarray}
with a Gaussian probability distribution with a width proportional to $dt$, namely with probability ($\int_{-\infty}^{\infty} p(\phi)d\phi=1$)
\begin{eqnarray}
p(\phi)&=&\frac{1}{\sqrt{2 \pi} \sigma}\cdot e^{-\frac{\phi^2}{2 \sigma^2}}\;,
\end{eqnarray}
where we choose explicitly the width $\sigma=\sqrt{\frac{\gamma}{(\sqrt{4\pi} r_C)^d}\cdot dt}$. Since we assume small $dt$ we can neglect safely the higher order terms and find for the density matrix at time $t+dt$
\begin{eqnarray}
\lefteqn{\hat{\rho}(t+dt)\;=\;\int_{-\infty}^{+\infty}d\phi\; p(\phi)\; \hat{U}(\phi)\cdot\hat{\rho}(t)\cdot \hat{U}^\dagger(\phi)}\nonumber\\
&=&   \int_{-\infty}^{+\infty}d\phi\; p(\phi)\; \left\lbrace\hat{\rho}(t)- \frac{\phi^2}{2}\left( \{\hat{G}^2,\hat{\rho}(t)\}-2\,\hat{G}\,\rho(t)\,\hat{G}\right)\right\rbrace\nonumber\\
&=& \hat{\rho}(t)- \frac{\sigma^2}{2}\left( \{\hat{G}^2,\hat{\rho}(t)\}-2\,\hat{G}\,\hat{\rho}(t)\;\hat{G}\right)\;.
\end{eqnarray}
This differential equation is equivalent to the one in the Lindblad form with $\hat{L}_1$ if we choose for $\hat{G}=\sum_i \frac{m_i}{m_0} |M_i\rangle\langle M_i|$ which is just the flavor part of our collapse operators. Even though we formally arrive at the same formulae, let us stress that in this case no real collapse is assumed, in particular the spatial part of the wave function played no role. Moreover, the dependence on the ``decay rate'' $\Gamma_i^{CSL}$ is not generated by the dynamics, but introduced by hand. However, it explains why the interference term in the flavor oscillation probabilities depends on $(\Delta m)^2$, this is a general feature of any random unitary noise with a Gaussian distribution.


\section{Summary and Outlook}\label{sec:Summary}
In this paper we have focused on two popular dynamical reduction models, QMUPL (Quantum Mechanics with Universal Position Localization) and mass-proportional CSL (Continuous Spontaneous Localization) model, and analyzed their effect on the neutral meson system. The beauty of these models is that they solve the measurement problem by introducing a physical mechanism for the collapse. In particular they assume that with a certain rate every quantum system undergoes a spontaneous localization in space. Taking these models seriously they have to also affect systems at higher energies, in particular, neutral mesons.

We have considered the two-state phenomenological Hamiltonian for a neutral meson system giving rise to flavor oscillations and assumed the collapse mechanism implied by dynamical reduction models. Since the collapse models assume the collapse to the spatial part of the state, we had to choose proper collapse operators relating the flavor space (where the oscillation takes place) with the spatial space. To calculate the effect of the collapse we have considered the (white) noise of the collapse models as a small perturbation by utilizing the Dyson series. The transition probabilities were calculated up to fourth perturbative order. This allowed to distinguish between exponential behavior (observed for the CSL model) and non-exponential behavior (observed for the QMUPL model), consequently giving insight into the physics of the noise field underlying the collapse mechanism.

Due to necessariness in solving the stochastic non-linear differential equations by perturbation series we have observed a dependence on the choice of the Heaviside function $\theta$ at time point $0$. This function turns up from integrating the correlations of two or more Wiener processes. Mathematically, the value is not well defined, it can be in the interval $\theta(0) \in [0,1]$. Only in the case of $\theta(0)=\frac{1}{2}$ we find that the lifetime state evolves independently of the collapse in spatial part. Indeed, having a closer view at the computations we observe that the value of the Heaviside function at time zero matters only if we evaluate amplitudes connecting different orders in the expansion. Consequently, the effect is due to the ``virtual'' propagation of the states and hence portrays a kind of consistency within the calculus. Since $\theta(0)=\frac{1}{2}$ physically implies that we have a time symmetric case of the two Wiener processes in the two dual spaces (ket/bra). This means that only in this case ($\theta(0)= \frac{1}{2}$) the norm of the state under investigation is conserved. Differently stated, the effect of the spontaneous collapse appears only in the interference terms of the flavor oscillation, not in the decaying parts.

Interestingly, any value $\theta(0)\not= \frac{1}{2}$ leads to a dependence on the absolute masses (energies) of the eigenstates of the time evolution, which does not show up in the standard quantum approach. In the CSL model they would not be directly measurable since they would effectively contribute to the decay constants of the standard quantum approach. On the other hand, the effect of the QMUPL model is in principle observable due to its non-exponential behavior. Since this deviation has not (yet) been observed, experiments provide upper bounds on the absolute masses of the lifetime states.

In a further step we analyzed whether the spontaneous localization could be considered as the only source of the decay in the neutral meson dynamics. We related the measured decay constants with the absolute masses appearing due to the nonlinear and stochastic modifications. The first observation was that for this identification the sign of the mass difference matters, i.e. if the longer-lived state corresponds to the more massive state or the lighter one. Experiments for K-mesons favor the first relation. This is in strong contrast to the philosophy of collapse models where a more massive system should localize faster in order to solve the measurement problem. To obtain still positive absolute masses one needs to identify the strength of the generation of a say heavy mass eigenstate with the lower mass. Doing so, we can deduce the absolute masses. Via that the decay mechanism of neutral mesons can be fully described. In turn we can also use them to predict values of the collapse rate (Fig.~\ref{figure1}). This rate is computed solely by the input parameter of the mass difference and the two decay rates in dependence on the value for $\theta(0)$. The range is in the expected region proposed by Adler or Ghirardi-Rimini-Weber except for the K-meson system.

In order to obtain a different insight into the physics behind the collapse models we also defined a master equation within standard quantum mechanics that leads to the same probabilities. Here we extended the Hilbert space to include the decay components as developed in Ref.~\cite{BGHOpenSystem}. Then a Gorini-Kossakowski-Lindblad-Sudarshan master equation with Gaussian noise proportional to the masses does the job. This illustrates the dependence of the damping on the squared mass difference since it is a general feature of systems undergoing a random Gaussian distributed unitary noise.

%
%
%
%

The strongest bounds on deviations from the expected quantum mechanical behaviors come from experiments with entangled mesons; or may be the case if the tiny $\mathbbm{CP}$ violating effects are taken into account. Therefore, it would be necessary to extend our computations to these cases. Moreover, our computations are performed for a white noise scenario. Colored noise should change the dynamics considerably and in turn allow to limit the collapse rate.

\vspace{0.5cm}
\textbf{Acknowledgment:}\\
The authors gratefully thank Sandro Donadi (Universit\`{a} degli studi di Trieste) for fruitful discussions and the Austrian Science Fund (FWF-P26783).
\vspace{0.5cm}

\appendix
\section{Computations for the QMUPL model}\label{sec:QMUPL_Computations}
\subsection{Transition probabilities for the mass eigenstates}

We start the computations of the transition probabilities for the QMUPL model with the $1$-dimensional case. The five terms (\ref{QMUPL_AmpZeroOrder})--(\ref{QMUPL_AmpFourthOrder}) form the transition amplitude up to fourth order of the Dyson series which we calculate here. Inserting the definition~(\ref{PerturbationQMUPL}) for the $\hat{N}_{QMUPL}(t)$ operator ($1$-dimensional case) and calculating the flavor part of matrix elements, we obtain the following expression for the components up to $n$-th order of the transition amplitudes
\begin{align}
\label{QMUPL_ComputationsAmp}
 & T^{(n)}_{\mu\nu}(p_f, p_i, \alpha ; t) = e^{-im_{\mu}t} F^{(n)}(t)  \Bigl( i \sqrt{\lambda} \frac{m_{\mu}}{m_0} \Bigr)^n \\
 & \nonumber \;\;\;\;\; \cdot \langle p_f | \hat{q}^n| p_i, \alpha \rangle\; \delta_{\mu\nu},
\end{align}
where
\begin{align}
 & \nonumber F^{(0)}(t_0) = 1\qquad\textrm{for all}\quad t_0, \\
 & \nonumber F^{(n)}(t_0) = \int_0^{t_0} dt_1 ... \int_0^{t_{n-1}} dt_n \prod_{j=1}^n w(t_j)\;. 
\end{align}
The transition amplitudes derive to
\begin{eqnarray}
\langle p_f | \hat{q}^n| p_i, \alpha \rangle &=& \sqrt{2\sqrt{\alpha\pi}}\; e^{-\frac{\alpha}{2}(p_f - p_i)^2}\cdot \zeta(n)
\end{eqnarray}
with
\begin{eqnarray*}
\zeta(0)&=&1,\\
\zeta(1)&=& (-i)\;\alpha\;(p_f - p_i),\\
\zeta(2)&=& \alpha\;(1-\alpha(p_f - p_i)^2),\\
\zeta(3)&=& (-i)\alpha^2(3 (p_f - p_i)-\alpha (p_f - p_i)^3),\\
\zeta(4)&=& \alpha^2 (3- 6 \alpha (p_f - p_i)^2+\alpha^2 (p_f - p_i)^4)\;.
\end{eqnarray*}

The transition probabilities up to second order in time $t$ are
\begin{align}\label{ComponentsProbQMUPL}
 & P_{M_{\mu} \rightarrow M_{\nu}}(\alpha ; t) = P^{(0)}_{M_{\mu} \rightarrow M_{\nu}}(\alpha ; t) + P^{(1)}_{M_{\mu} \rightarrow M_{\nu}}(\alpha ; t) \\
 & \nonumber \;\;\;\;\; + P^{(2)}_{M_{\mu} \rightarrow M_{\nu}}(\alpha ; t),
\end{align}
which decompose in terms of the transition amplitudes to
\begin{widetext}
\begin{subequations}
\begin{align}
 P^{(0)}_{M_{\mu} \rightarrow M_{\nu}}(\alpha ; t) & = \frac{1}{2\pi} \int dp_f \mathbb{E} \Bigl[T_{\mu\nu}^{(0)}(p_f, p_i, \alpha ; t) T_{\mu\nu}^{(0)*}(p_f, p_i, \alpha ; t)\Bigr], \\
 P^{(1)}_{M_{\mu} \rightarrow M_{\nu}}(\alpha ; t) & = \frac{1}{2\pi} \int dp_f \mathbb{E} \Bigl[T_{\mu\nu}^{(0)}(p_f, p_i, \alpha ; t) T_{\mu\nu}^{(2)*}(p_f, p_i, \alpha ; t) + T_{\mu\nu}^{(2)}(p_f, p_i, \alpha ; t) T_{\mu\nu}^{(0)*}(p_f, p_i, \alpha ; t) \\
 & \nonumber \;\;\;\;\; + T_{\mu\nu}^{(1)}(p_f, p_i, \alpha ; t) T_{\mu\nu}^{(1)*}(p_f, p_i, \alpha ; t)\Bigr], \\
 P^{(2)}_{M_{\mu} \rightarrow M_{\nu}}(\alpha ; t) & = \frac{1}{2\pi} \int dp_f \mathbb{E} \Bigl[T_{\mu\nu}^{(0)}(p_f, p_i, \alpha ; t) T_{\mu\nu}^{(4)*}(p_f, p_i, \alpha ; t) + T_{\mu\nu}^{(4)}(p_f, p_i, \alpha ; t) T_{\mu\nu}^{(0)*}(p_f, p_i, \alpha ; t) \\
 & \nonumber \;\;\;\;\; + T_{\mu\nu}^{(1)}(p_f, p_i, \alpha ; t) T_{\mu\nu}^{(3)*}(p_f, p_i, \alpha ; t) + T_{\mu\nu}^{(3)}(p_f, p_i, \alpha ; t) T_{\mu\nu}^{(1)*}(p_f, p_i, \alpha ; t) \\
 & \nonumber \;\;\;\;\; + T_{\mu\nu}^{(2)}(p_f, p_i, \alpha ; t) T_{\mu\nu}^{(2)*}(p_f, p_i, \alpha ; t)\Bigr].
\end{align}
\end{subequations}
\end{widetext}
The first term gives
\begin{align}\label{ZeroComponentProbQMUPL}
& P^{(0)}_{M_{\mu}\rightarrow M_{\nu}}(\alpha ; t) \\
& \nonumber = \frac{2\sqrt{\alpha\pi}}{2\pi}\int dp_f\;  e^{-\alpha(p_f - p_i)^2} \delta_{\mu\nu} = \delta_{\mu\nu}\;.
\end{align}
For the first order in time we need
\begin{eqnarray*}
&&\frac{1}{2\pi} \int dp_f \;\mathbb{E} \Bigl[ T^{(0)*}_{\mu\nu}(p_f, p_i, \alpha ; t) T^{(2)}_{\mu\nu}(p_f, p_i, \alpha ; t) \\
&&\qquad  + \; T^{(2)*}_{\mu\nu}(p_f, p_i, \alpha ; t) T^{(0)}_{\mu\nu}(p_f, p_i, \alpha ; t) \Bigr] \\
&=&  -2\delta_{\mu\nu} \frac{\lambda m_{\mu}^2}{m_0^2}\; \int_0^t dt_1 \int_0^{t_1} dt_2 \mathbb{E}[w(t_1) w(t_2)] \\
&& \;\;\;\;\; \cdot \frac{2\alpha\sqrt{\alpha\pi}}{2\pi}\int dp_f \Bigl[ 1 - \alpha (p_f - p_i)^2 \Bigr] e^{-\alpha(p_f - p_i)^2} \\
&=& -\delta_{\mu\nu}\; (\alpha\lambda) \frac{ m_{\mu}^2}{m_0^2} \Bigl(1-\theta(0)\Bigr)\cdot t\;,
\end{eqnarray*}
where the computation of the two-point correlation function $\int_0^t dt_1 \int_0^{t_1} dt_2 \mathbb{E}[w(t_1) w(t_2)]$ is explicitly derived in the Appendix~\ref{sec:TimeIntegrals}.

The second term derives to
\begin{align}
 & \nonumber \frac{1}{2\pi} \int dp_f \mathbb{E} \Bigl[ T^{(1)*}_{\mu\nu}(p_f, p_i, \alpha ; t) T^{(1)}_{\mu\nu}(p_f, p_i, \alpha ; t) \Bigr] \\
 & \nonumber = \delta_{\mu\nu} \frac{\lambda m_{\mu}^2}{m_0^2}  \int_0^t dt_1 \int_0^{t} dt_2 \mathbb{E}[w(t_1) w(t_2)] \\
 & \nonumber \;\;\;\;\; \cdot \frac{2\alpha^2 \sqrt{\alpha\pi}}{2\pi}\int dp_f (p_f-p_i)^2 e^{-\alpha(p_f - p_i)^2} \\
 & \nonumber = \delta_{\mu\nu}\frac{\alpha}{2} \frac{\lambda m_{\mu}^2}{m_0^2}\cdot t\;,
\end{align}
where the two-point correlation function $\int_0^t dt_1 \int_0^{t} dt_2 \mathbb{E}[w(t_1) w(t_2)]$ is derived in the Appendix~\ref{sec:TimeIntegrals} (note the difference in the integration limits).

Consequently, the transition probabilities in first order in time $t$ result in
\begin{align}\label{FirstComponentProbQMUPL}
& P^{(1)}_{M_{\mu}\rightarrow M_{\nu}}(\alpha ; t) = -\delta_{\mu\nu} \frac{\alpha}{2} \frac{\lambda m_{\mu}^2}{m_0^2} (1-2\theta(0))\cdot t.
\end{align}

To obtain the solution in the second order in time $t$ we have to compute the five components, $T^{(0)*}T^{(4)}$, $T^{(4)*}T^{(0)}$, $T^{(1)*}T^{(3)}$, $T^{(3)*}T^{(1)}$ and $T^{(2)*}T^{(2)}$, where we have for the first time to evaluate a four point function in the noise which is done in detail in Appendix~\ref{sec:TimeIntegrals}. We compute
\begin{align}
& \nonumber \frac{1}{2\pi} \int dp_f \mathbb{E} \Bigl[ T^{(0)*}_{\mu\nu}(p_f, p_i, \alpha ; t) T^{(4)}_{\mu\nu}(p_f, p_i, \alpha ; t) \\
& \nonumber + T^{(0)*}_{\mu\nu}(p_f, p_i, \alpha ; t) T^{(4)}_{\mu\nu}(p_f, p_i, \alpha ; t) \Bigr] \\
& \nonumber = 2\delta_{\mu\nu} \frac{\lambda^2 m_{\mu}^4}{m_0^4} C^{(2)}_{4,0} (t) \\
& \nonumber \;\;\;\;\; \cdot \frac{2\alpha^2\sqrt{\alpha\pi}}{2\pi}\int dp_f \Bigl[ 3 - 6\alpha (p_f - p_i)^2 \\
& \nonumber \;\;\;\;\; + \alpha^2 (p_f - p_i)^4 \Bigr] e^{-\alpha(p_f - p_i)^2} \\
& \nonumber = \delta_{\mu\nu} \frac{3\alpha^2}{2} \frac{\lambda^2 m_{\mu}^4}{m_0^4} C^{(2)}_{4,0} (t) = \delta_{\mu\nu} \frac{3\alpha^2}{2} \frac{\lambda^2 m_{\mu}^4}{m_0^4} \cdot \frac{1}{2} \Bigl(1 - \theta(0) \Bigr)^2\cdot t^2,
\end{align}
and
\begin{align}
& \nonumber \frac{1}{2\pi} \int dp_f \mathbb{E} \Bigl[ T^{(1)*}_{\mu\nu}(p_f, p_i, \alpha ; t) T^{(3)}_{\mu\nu}(p_f, p_i, \alpha ; t) \\
& \nonumber \;\;\;\;\; + T^{(3)*}_{\mu\nu}(p_f, p_i, \alpha ; t) T^{(1)}_{\mu\nu}(p_f, p_i, \alpha ; t) \Bigr] \\
& \nonumber = -2\delta_{\mu\nu} \frac{\lambda^2 m_{\mu}^4}{m_0^4} C^{(2)}_{3,1} (t) \\
& \nonumber \;\;\;\;\; \cdot \frac{2\alpha^3\sqrt{\alpha\pi}}{2\pi}\int dp_f \Bigl[ 3 - \alpha(p_f-p_i)^2 \Bigr] (p_f-p_i)^2 \\
& \nonumber \;\;\;\;\; \cdot e^{-\alpha(p_f - p_i)^2} \\
& \nonumber = -\delta_{\mu\nu} \frac{3\alpha^2}{2} \frac{\lambda^2 m_{\mu}^4}{m_0^4} C^{(2)}_{3,1} (t) \\
& \nonumber = -\delta_{\mu\nu} \frac{3\alpha^2}{2} \frac{\lambda^2 m_{\mu}^4}{m_0^4} \cdot \Bigl( 1 - \theta(0) \Bigr)\cdot t^2,
\end{align}
and
\begin{align}
& \nonumber \frac{1}{2\pi} \int dp_f \mathbb{E} \Bigl[ T^{(2)*}_{\mu\nu}(p_f, p_i, \alpha ; t) T^{(2)}_{\mu\nu}(p_f, p_i, \alpha ; t) \Bigr] \\
& \nonumber = \delta_{\mu\nu} \frac{\lambda^2 m_{\mu}^4}{m_0^4} C^{(2)}_{2,2} (t) \\
& \nonumber \;\;\;\;\; \cdot \frac{2\alpha^2\sqrt{\alpha\pi}}{2\pi}\int dp_f \Bigl[ 1 - \alpha(p_f-p_i)^2 \Bigr]^2 e^{-\alpha(p_f - p_i)^2} \\
& \nonumber = \delta_{\mu\nu} \frac{3\alpha^2}{4} \frac{\lambda^2 m_{\mu}^4}{m_0^4} C^{(2)}_{2,2} (t) \\
& \nonumber = \delta_{\mu\nu} \frac{3\alpha^2}{4} \frac{\lambda^2 m_{\mu}^4}{m_0^4} \cdot \Bigl( \Bigl( 1 - \theta(0) \Bigr)^2 + \frac{1}{2} \Bigr)\cdot t^2.
\end{align}
where $C^{(2)}_{4,0} (t)$, $C^{(2)}_{3,1} (t)$ and $C^{(2)}_{2,2} (t)$ correspond to the integrals of the $4$-point correlation functions of the noise field, which are calculated in Appendix~\ref{sec:TimeIntegrals}.

Summing up, we obtain the transition probabilities in second order time $t$
\begin{align}\label{SecondComponentProbQMUPL}
& P^{(2)}_{M_{\mu}\rightarrow M_{\nu}} (\alpha; t) \\
& \nonumber = \delta_{\mu\nu} \frac{3\alpha^2}{4} \frac{\lambda^2 m_{\mu}^4}{m_0^4} \Bigl(2\theta(0) (\theta(0) - 1 ) + \frac{1}{2} \Bigr)\cdot t^2 \\
& \nonumber = \delta_{\mu\nu} \frac{3\alpha^2}{8} \frac{\lambda^2 m_{\mu}^4}{m_0^4} \Bigl(1 - 2\theta(0) \Bigr)^2 \cdot t^2.
\end{align}
Finally, collecting all the terms~(\ref{ZeroComponentProbQMUPL})--(\ref{SecondComponentProbQMUPL}), we obtain the transition probabilities for mass eigenstates up to second order in time $t$
\begin{align}
& P_{M_{\mu}\rightarrow M_{\nu}} (\alpha; t) = \delta_{\mu\nu}\Bigl[ 1 - \frac{\alpha}{2} \frac{\lambda m_{\mu}^2}{m_0^2} \Bigl(1-2\theta(0) \Bigr) t \\
& \nonumber \;\;\;\;\; + \frac{3\alpha^2}{8} \frac{\lambda^2 m_{\mu}^4}{m_0^4} \Bigl(1 - 2\theta(0) \Bigr)^2 t^2 \Bigr].
\end{align}

\subsection{$d$-dimensional case}\label{sec:QMUPL_MultiDim}
In the case of $d$-dimensional space the components (\ref{QMUPL_AmpZeroOrder})--(\ref{QMUPL_AmpFourthOrder}) of transition amplitudes have to be generalized in the following way
\begin{align}\label{QMUPLMassStatesAmpMulti}
 & T^{(n)}_{\mu\nu}(\mathbf{p}_f, \mathbf{p}_i, \alpha ; t) \\
 & \nonumber = e^{-im_{\mu}t}\, \tilde{F}^{(n)}(\mathbf{p}_f, \mathbf{p}_i, \alpha ; t)\, \Bigl( i \sqrt{\lambda} \frac{m_{\mu}}{m_0} \Bigr)^n \delta_{\mu\nu},
\end{align}
where
\begin{align}
 & \nonumber \tilde{F}^{(0)}(\mathbf{p}_f, \mathbf{p}_i, \alpha ; t_0) = \langle \mathbf{p}_f | \mathbf{p}_i, \alpha \rangle, \\
 & \nonumber \tilde{F}^{(n)}(\mathbf{p}_f, \mathbf{p}_i, \alpha ; t_0) = \int_0^{t_0} dt_1 ... \int_0^{t_{n-1}} dt_n \\
 & \nonumber \;\;\;\;\; \cdot \langle \mathbf{p}_f | \prod_{j=1}^n (\hat{\mathbf{q}} \cdot \mathbf{w}(t_j) ) | \mathbf{p}_i, \alpha \rangle.
\end{align}
Here one can think of basically two different ways the noise would act onto the system. Either a factorization in any of the possible dimensions happens and contributes to the first order in time, or a factorization of the wave function has to occur in all dimensions simultaneously. The second one seems to be less natural to assume. Since we assume white noise and an initial Gaussian wave function in all dimensions, however, integrals give the same value and the only difference is how often the integral occurs. Therefore, we stick to the first case.

Explicitly, we find
\begin{widetext}
\begin{align}
 & \nonumber \tilde{F}_0(\mathbf{p}_f, \mathbf{p}_i, \alpha ; t) = \Bigl(2\sqrt{\alpha\pi}\Bigr)^{d/2} e^{-\frac{\alpha}{2}(\mathbf{p}_f - \mathbf{p}_i)^2}, \\
 & \nonumber \tilde{F}_1(\mathbf{p}_f, \mathbf{p}_i, \alpha ; t) = -i \cdot \Bigl(2\sqrt{\alpha\pi}\Bigr)^{d/2} \alpha \int\limits_0^t dt_1  \Bigl( (\mathbf{p}_f-\mathbf{p}_i) \cdot \mathbf{w}(t_1) \Bigr) e^{-\frac{\alpha}{2}(\mathbf{p}_f - \mathbf{p}_i)^2}, \\
 & \nonumber \tilde{F}_2(\mathbf{p}_f, \mathbf{p}_i, \alpha ; t) = \Bigl(2\sqrt{\alpha\pi}\Bigr)^{d/2}\alpha\int\limits_0^t dt_1 \int\limits_0^{t_1} dt_2 \Biggl[ (\mathbf{w}(t_1) \cdot \mathbf{w}(t_2) ) - \alpha \Bigl((\mathbf{p}_f - \mathbf{p}_i) \cdot \mathbf{w}(t_1) \Bigr)\Bigl((\mathbf{p}_f - \mathbf{p}_i) \cdot \mathbf{w}(t_2) \Bigr) \Biggr] e^{-\frac{\alpha}{2}(\mathbf{p}_f - \mathbf{p}_i)^2}, \\
 & \nonumber \tilde{F}_3(\mathbf{p}_f, \mathbf{p}_i, \alpha ; t) = -i \cdot \Bigl(2\sqrt{\alpha\pi}\Bigr)^{d/2}\alpha^2 \int\limits_0^t dt_1 \int\limits_0^{t_1} dt_2 \int\limits_0^{t_2} dt_3 \Biggl[ \Bigl((\mathbf{p}_f-\mathbf{p}_i) \cdot \mathbf{w}(t_1) \Bigr)\Bigl(\mathbf{w}(t_2)\cdot \mathbf{w}(t_3)\Bigr) \\
 & \nonumber \;\;\;\;\; + \Bigl((\mathbf{p}_f-\mathbf{p}_i) \cdot \mathbf{w}(t_2) \Bigr)\Bigl(\mathbf{w}(t_1)\cdot \mathbf{w}(t_3)\Bigr) + \Bigl((\mathbf{p}_f-\mathbf{p}_i) \cdot \mathbf{w}(t_3) \Bigr)\Bigl(\mathbf{w}(t_1)\cdot \mathbf{w}(t_2)\Bigr) \\
 & \nonumber \;\;\;\;\;  - \alpha \Bigl( (\mathbf{p}_f-\mathbf{p}_i) \cdot \mathbf{w}(t_1) \Bigr)\Bigl( (\mathbf{p}_f-\mathbf{p}_i) \cdot \mathbf{w}(t_2) \Bigr)\Bigl( (\mathbf{p}_f-\mathbf{p}_i) \cdot \mathbf{w}(t_3) \Bigr) \Biggr] e^{-\frac{\alpha}{2}(\mathbf{p}_f - \mathbf{p}_i)^2}, \\
 & \nonumber \tilde{F}_4(\mathbf{p}_f, \mathbf{p}_i, \alpha ; t) = \Bigl(2\sqrt{\alpha\pi}\Bigr)^{d/2}\alpha^2 \int\limits_0^t dt_1 \int\limits_0^{t_1} dt_2 \int\limits_0^{t_2} dt_3\int\limits_0^{t_3} dt_4 \Biggl[ \Bigl(\mathbf{w}(t_1)\cdot \mathbf{w}(t_2)\Bigr) \Bigl(\mathbf{w}(t_3)\cdot \mathbf{w}(t_4)\Bigr) + \Bigl(\mathbf{w}(t_1)\cdot \mathbf{w}(t_3)\Bigr) \Bigl(\mathbf{w}(t_2)\cdot \mathbf{w}(t_4)\Bigr) \\
 & \nonumber \;\;\;\;\; + \Bigl(\mathbf{w}(t_1)\cdot \mathbf{w}(t_4)\Bigr) \Bigl(\mathbf{w}(t_2)\cdot \mathbf{w}(t_3)\Bigr) - \alpha \Bigl((\mathbf{p}_f - \mathbf{p}_i) \cdot \mathbf{w}(t_1) \Bigr)\Bigl((\mathbf{p}_f - \mathbf{p}_i) \cdot \mathbf{w}(t_2) \Bigr)\Bigl(\mathbf{w}(t_3) \cdot \mathbf{w}(t_4) \Bigr) \\
 & \nonumber \;\;\;\;\; - \alpha \Bigl((\mathbf{p}_f - \mathbf{p}_i) \cdot \mathbf{w}(t_1) \Bigr)\Bigl((\mathbf{p}_f - \mathbf{p}_i) \cdot \mathbf{w}(t_3) \Bigr)\Bigl(\mathbf{w}(t_2) \cdot \mathbf{w}(t_4) \Bigr) - \alpha \Bigl((\mathbf{p}_f - \mathbf{p}_i) \cdot \mathbf{w}(t_1) \Bigr)\Bigl((\mathbf{p}_f - \mathbf{p}_i) \cdot \mathbf{w}(t_4) \Bigr)\Bigl(\mathbf{w}(t_2) \cdot \mathbf{w}(t_3) \Bigr) \\
 & \nonumber \;\;\;\;\; - \alpha \Bigl((\mathbf{p}_f - \mathbf{p}_i) \cdot \mathbf{w}(t_2) \Bigr)\Bigl((\mathbf{p}_f - \mathbf{p}_i) \cdot \mathbf{w}(t_3) \Bigr)\Bigl(\mathbf{w}(t_1) \cdot \mathbf{w}(t_4) \Bigr) - \alpha \Bigl((\mathbf{p}_f - \mathbf{p}_i) \cdot \mathbf{w}(t_2) \Bigr)\Bigl((\mathbf{p}_f - \mathbf{p}_i) \cdot \mathbf{w}(t_4) \Bigr)\Bigl(\mathbf{w}(t_1) \cdot \mathbf{w}(t_3) \Bigr) \\
 & \nonumber \;\;\;\;\; - \alpha \Bigl((\mathbf{p}_f - \mathbf{p}_i) \cdot \mathbf{w}(t_3) \Bigr)\Bigl((\mathbf{p}_f - \mathbf{p}_i) \cdot \mathbf{w}(t_4) \Bigr)\Bigl(\mathbf{w}(t_1) \cdot \mathbf{w}(t_2) \Bigr) \\
 & \nonumber \;\;\;\;\; + \alpha^2 \Bigl((\mathbf{p}_f - \mathbf{p}_i) \cdot \mathbf{w}(t_1) \Bigr)\Bigl((\mathbf{p}_f - \mathbf{p}_i) \cdot \mathbf{w}(t_2) \Bigr)\Bigl((\mathbf{p}_f - \mathbf{p}_i) \cdot \mathbf{w}(t_3) \Bigr)\Bigl((\mathbf{p}_f - \mathbf{p}_i) \cdot \mathbf{w}(t_4) \Bigr) \Biggr] e^{-\frac{\alpha}{2}(\mathbf{p}_f - \mathbf{p}_i)^2}.
\end{align}
\end{widetext}
\begin{widetext}
and herewith the probabilities
\begin{align}
 & \nonumber P^{(0)}_{M_{\mu}\rightarrow M_{\nu}}(\alpha ; t) = \delta_{\mu\nu}, \\
 & \nonumber P^{(1)}_{M_{\mu}\rightarrow M_{\nu}}(\alpha ; t) = -\delta_{\mu\nu} \frac{\alpha}{2} \frac{\lambda m_{\mu}^2}{m_0^2} \Bigl(1 - 2\theta(0) \Bigr) t, \\
 & \nonumber  P^{(2)}_{M_{\mu}\rightarrow M_{\nu}}(\alpha ; t) = \delta_{\mu\nu} \frac{3\alpha^2}{4} \frac{\lambda^2 m_{\mu}^4}{m_0^4} \Bigl(2\theta(0) (\theta(0) - 1 ) + \frac{1}{2} \Bigr) t^2\;,
\end{align}
\end{widetext}
which are identical to the ones of the $1$-dimensional case and, consequently, lead to the same transition probabilities.

\subsection{Transition probabilities for the flavor states}
Transition amplitude for a flavor state can be expanded in the following way:
\begin{align}
& \nonumber T_{M^0 \rightarrow M^0/\bar{M}^0} (\mathbf{p}_f, \mathbf{p}_i, \alpha; t) = \langle M^0/\bar{M}^0, \mathbf{p}_f | M^0(t), \mathbf{p}_i, \alpha \rangle \\
& \nonumber = \sum_{\mu, \nu} \alpha_{\mu} \beta^*_{\nu}\langle M_\nu, \mathbf{p}_f | M_\mu (t), \mathbf{p}_i, \alpha \rangle \\
& \nonumber = \sum_{\mu, \nu} \alpha_{\mu} \beta^*_{\nu} T_{\mu\nu}(\mathbf{p}_f, \mathbf{p}_i, \alpha; t),
\end{align}
where $\mu,\nu = H,L$ and $\alpha_H = \alpha_L = \beta_H = \frac{1}{\sqrt{2}}$, $\beta_L = \pm \frac{1}{\sqrt{2}}$ (plus sign refers to a meson, minus sign refers to an antimeson). In the same manner, transition probability for a flavor state can be defined as
\begin{align}
& P_{M^0 \rightarrow M^0/\bar{M}^0}(\alpha; t) = \sum_{\mu, \nu, \mu', \nu'} \alpha_{\mu} \beta^*_{\nu} \alpha^*_{\mu'} \beta_{\nu'} \\
& \nonumber \;\;\;\;\; \cdot \frac{1}{(2\pi)^d} \int d\mathbf{p}_f \mathbb{E} [T_{\mu\nu}(\mathbf{p}_f, \mathbf{p}_i, \alpha; t) T^*_{\mu'\nu'}(\mathbf{p}_f, \mathbf{p}_i, \alpha; t)] \\
& \nonumber \equiv \sum_{\mu, \nu, \mu', \nu'} \alpha_{\mu} \beta^*_{\nu} \alpha^*_{\mu'} \beta_{\nu'} P_{\mu \nu \mu' \nu'}(\alpha; t),
\end{align}
Furthermore, since each transition amplitude $T_{\mu\nu}(\mathbf{p}_f, \mathbf{p}_i, \alpha; t)$ contains a Kronecker delta $\delta_{\mu\nu}$, as can be seen from~(\ref{QMUPL_ComputationsAmp}) and (\ref{QMUPLMassStatesAmpMulti}), we can leave just one index in an amplitude and correspondingly two indexes in probabilities $P_{\mu \nu \mu' \nu'}(\alpha; t)$
\begin{align}
& P_{M^0 \rightarrow M^0/\bar{M}^0}(\alpha; t) = \sum_{\mu, \mu'} \alpha_{\mu} \beta^*_{\mu} \alpha^*_{\mu'} \beta_{\mu'} P_{\mu \mu'}(\alpha; t) \\
& \nonumber = \frac{1}{4} (P_{HH}(\alpha; t) \pm P_{HL}(\alpha; t) \pm P_{LH}(\alpha; t) + P_{LL}(\alpha; t)),
\end{align}
Using the transition probabilities which were calculated above we obtain the terms for the transition probability, with same indexes $P_{aa}$ and different ones $P_{ab}$
\begin{align}
& P_{aa}(\alpha; t) = 1 - \frac{\alpha}{2} \frac{\lambda m_a^2}{m_0^2} \Bigl( 1 - 2\theta(0) \Bigr) t \\
& \nonumber \;\;\;\;\; + \frac{3\alpha^2}{8} \frac{\lambda^2 m_a^4}{m_0^4} \Bigl(1 - 2\theta(0) \Bigr)^2 t^2, \\
& P_{ab}(\alpha; t) = e^{-i(m_a - m_b)t} \\
& \nonumber \;\;\;\;\; \cdot \Biggl\{ 1 - \frac{\alpha}{2} \frac{\lambda}{m_0^2} \Bigl( (m_a^2 + m_b^2)\Bigl( 1 -\theta(0) \Bigr) - m_a m_b\Bigr) t \\
& \nonumber \;\;\;\;\; + \frac{3\alpha^2}{8} \frac{\lambda^2}{m_0^4} \Bigl[(m_a^4 + m_b^4) \Bigl(1 - \theta(0)\Bigr)^2 \\
& \nonumber \;\;\;\;\; - 2(m_a^3 m_b + m_a m_b^3) \Bigl(1 - \theta(0)\Bigr) \\
& \nonumber \;\;\;\;\; + 2m_a^2 m_b^2 \Bigl(\Bigl(1 - \theta(0)\Bigr)^2 + \frac{1}{2}\Bigr) \Bigr] t^2 \Biggr\}.
\end{align}
Putting the terms together we finally obtain the transition probability for the flavor states
\begin{widetext}
\begin{align}
 & P_{M^0 \rightarrow M^0/\bar{M}^0}(\alpha; t) = \frac{1}{2} \Biggl\{ 1 - \frac{\alpha}{4} \frac{\lambda (m_H^2 + m_L^2)}{m_0^2} \Bigl( 1 - 2\theta(0)\Bigr) \cdot t + \frac{3\alpha^2}{16} \frac{\lambda^2 (m_H^4 + m_L^4)}{m_0^4} \Bigl(1 - 2\theta(0) \Bigr)^2 \cdot t^2 \\
 & \nonumber \;\;\;\;\; \pm \Biggl[1 - \frac{1}{2} \frac{\lambda \alpha}{m_0^2} \Bigl( (m_H^2 + m_L^2)\Bigl(1 - \theta(0)\Bigr) - m_H m_L\Bigr) \cdot t + \frac{3}{8} \frac{\lambda^2 \alpha^2}{m_0^4} \Bigl( (m_H^4 + m_L^4)\Bigl(1 - \theta(0)\Bigr)^2  \\
 & \nonumber \;\;\;\;\; - 2\,m_H m_L (m_H^2 + m_L^2)\Bigl(1 - \theta(0)\Bigr) + 2m_H^2 m_L^2 \Bigl(\Bigl(1 - \theta(0)\Bigr)^2 + \frac{1}{2} \Bigr) \Bigr) \cdot t^2 \Biggr]\cdot\cos\Bigl[(m_H - m_L)t\Bigr]\Biggr\}.
\end{align}

Taking the decay into account we obtain
\begin{align}
& P_{M^0 \rightarrow M^0/\bar{M}^0}(\alpha; t) = \frac{1}{4} \Biggl\{ e^{-\Gamma_H t}+e^{-\Gamma_L t} - \frac{1}{2} \frac{\lambda\alpha}{m_0^2} (m_H^2 e^{-\Gamma_H t}+ m_L^2 e^{-\Gamma_L t})\Bigl(1 - 2\theta(0)\Bigr) \cdot t \\
&\nonumber \;\;\;\;\; + \frac{3}{8} \frac{\lambda^2\alpha^2}{m_0^4} (m_H^4 e^{-\Gamma_H t}+ m_L^4e^{-\Gamma_L t})\Bigl(1 - 2\theta(0) \Bigr)^2 \cdot t^2 \pm 2\Biggl[1 - \frac{1}{2} \frac{\lambda\alpha}{m_0^2} \Bigl( (m_H^2 + m_L^2)\Bigl(1 - \theta(0)\Bigr) - m_H m_L\Bigr) \cdot t \\
& \nonumber \;\;\;\;\; + \frac{3}{8} \frac{\lambda^2\alpha^2}{m_0^4} \Bigl( (m_H^4 + m_L^4)\Bigl(1 - \theta(0)\Bigr)^2 - 2m_H m_L (m_H^2 + m_L^2)\Bigl(1 - \theta(0)\Bigr) + 2m_H^2 m_L^2 \Bigl(\Bigl(1 - \theta(0)\Bigr)^2 + \frac{1}{2} \Bigr) \Bigr) \cdot t^2 \Biggr] \\
& \nonumber \;\;\;\;\; \cdot\cos\Bigl[(m_H - m_L)t\Bigr] \cdot e^{-\frac{\Gamma_H+\Gamma_L}{2} t}\Biggr\}.
\end{align}
\end{widetext}

\section{Computations for the CSL model}\label{sec:CSL_Computations}
\subsection{Transition probabilities for mass eigenstates}

For the CSL model we also have five terms which form the transition amplitude up to fourth order of the Dyson series. Putting the expressions for the $\hat{N}_I$ operators in we obtain:
\begin{align}
 & \nonumber T^{(n)}_{\mu\nu}(\mathbf{p}_f, \mathbf{p}_i, \alpha ; t) = e^{-im_{\mu} t} (i\sqrt{\gamma})^n K_{\mu\nu}^{(n)} (\mathbf{p}_f, \mathbf{p}_i, \alpha ; t),
\end{align}
where
\begin{align}
& \nonumber K_{\mu\nu}^{(0)} (\mathbf{p}_f, \mathbf{p}_i, \alpha ; t_0) = \langle M_{\nu}, \mathbf{p}_f  | M_{\mu}, \mathbf{p}_i, \alpha \rangle,\\
& \nonumber K_{\mu\nu}^{(n)} (\mathbf{p}_f, \mathbf{p}_i, \alpha ; t_0) = \int\limits_0^{t_0} dt_1 \dots \int\limits_0^{t_{n-1}} dt_n \int d\mathbf{x}_1 \dots \int d\mathbf{x}_n \\
& \nonumber \;\;\;\;\; \cdot \langle M_{\nu}, \mathbf{p}_f | \prod\limits_{j=1}^n \Bigl(w(t_j, \mathbf{x}_j) \\
& \nonumber \;\;\;\;\; \cdot \sum_{k=H,L} \frac{m_k}{m_0} \hat{\psi}_I^{k\dagger}(t_j, \mathbf{x}_j) \hat{\psi}_I^{k}(t_j, \mathbf{x}_j)\Bigr) | M_{\mu}, \mathbf{p}_i, \alpha \rangle.
\end{align}
Accordingly, we will calculate the matrix elements in the same manner as done in~\cite{Donadi2013}. At first, we make an expansion of field operators into a superposition of plane waves
\begin{align}\label{PsiFourier}
& \hat{\psi}_I^k (t, x) = \frac{1}{\sqrt{L^d}}\sum_{\mathbf{q}} \hat{b}_{\mathbf{q}} e^{-i(E^{(k)}_q t - \mathbf{q}\cdot\mathbf{x})},
\end{align}
where the energy of a meson of mass $m_k$ and momentum $\mathbf{q}$ is taken in non-relativistic limit, $E^{(k)}_q = \sqrt{\mathbf{q}^2 + m_k^2} \approx m_k$. Here the system is assumed to be quantized in a box of size $L$ with using periodic boundary conditions. While calculating the transition amplitudes and probabilities we take the limit $L\rightarrow\infty$ and perform an integration by momentum $\frac{1}{\sqrt{L^d}}\sum_\mathbf{q} \rightarrow \frac{1}{\sqrt{(2\pi)^d}}\int d\mathbf{q}$.

Using the coordinate representation and calculating the matrix elements, we obtain components of the transition amplitudes in the following form
\begin{widetext}
\begin{subequations}
\begin{align}\label{ZerothComponentProbCSL}
& \nonumber K^{(0)}_{\mu\nu}(\mathbf{p}_f, \mathbf{p}_i ; t) = ( 2\sqrt{\alpha\pi} )^{d/2} e^{-\frac{\alpha}{2}(\mathbf{p}_f - \mathbf{p}_i)^2} \delta_{\mu\nu}, \\
& \nonumber K^{(1)}_{\mu\nu}(\mathbf{p}_f, \mathbf{p}_i; t) = \frac{m_\mu}{m_0} \Bigr[\Bigl( \frac{1}{\sqrt{\alpha\pi}} \Bigr)^{d/2} \int\limits_0^t dt_1 \int d\mathbf{x}_1 w(t_1, \mathbf{x}_1) \cdot e^{- i (\mathbf{p}_f - \mathbf{p}_i) \mathbf{x}_1 } e^{-\frac{\mathbf{x}_1^2}{2\alpha}} \Bigl] \delta_{\mu\nu}, \\
& \nonumber K^{(2)}_{\mu\nu}(\mathbf{p}_f, \mathbf{p}_i ; t) = \frac{m_\mu^2}{m_0^2} \frac{1}{(2\pi)^d} \int d\mathbf{q}_1 \; \Bigl[ \Bigl( \frac{1}{\sqrt{\alpha\pi}} \Bigr)^{d/2} \int\limits_0^t dt_1 \int\limits_0^{t_1} dt_2 \int\!\!\int d\mathbf{x}_1 d\mathbf{x}_2 \\
& \nonumber \;\;\;\;\; \cdot w(t_1,\mathbf{x}_1) w(t_2,\mathbf{x}_2) \cdot e^{- i (\mathbf{p}_f - \mathbf{q}) \mathbf{x}_1 } e^{- i (\mathbf{q} - \mathbf{p}_i) \mathbf{x}_2 } e^{-\frac{\mathbf{x}_2^2}{2\alpha}} \Bigl]\delta_{\mu\nu}, \\
& \nonumber K^{(3)}_{\mu\nu}(\mathbf{p}_f, \mathbf{p}_i ; t) = \frac{m_\mu^3}{m_0^3} \frac{1}{(2\pi)^{2d}} \int\!\!\int d\mathbf{q}_1 d\mathbf{q}_2 \; \Bigl[ \Bigl( \frac{1}{\sqrt{\alpha\pi}} \Bigr)^{d/2} \int\limits_0^t dt_1 \int\limits_0^{t_1} dt_2 \int\limits_0^{t_2} dt_3 \int\!\!\int\!\!\int d\mathbf{x}_1 d\mathbf{x}_2 d\mathbf{x}_3 \\
& \nonumber \;\;\;\;\; \cdot w(t_1,\mathbf{x}_1) w(t_2,\mathbf{x}_2) w(t_3,\mathbf{x}_3) \cdot e^{- i (\mathbf{p}_f - \mathbf{q}_1) \mathbf{x}_1 } e^{- i (\mathbf{q}_1 - \mathbf{q}_2) \mathbf{x}_2 } e^{- i (\mathbf{q}_2 - \mathbf{p}_i) \mathbf{x}_3 } e^{-\frac{\mathbf{x}_3^2}{2\alpha}} \Bigl]\delta_{\mu\nu}, \\
& \nonumber K^{(4)}_{\mu\nu}(\mathbf{p}_f, \mathbf{p}_i, \alpha ; t) = \frac{m_\mu^4}{m_0^4} \frac{1}{(2\pi)^{3d}} \int\!\!\int\!\!\int d\mathbf{q}_1 d\mathbf{q}_2 d\mathbf{q}_3 \; \Bigl[ \Bigl( \frac{1}{\sqrt{\alpha\pi}} \Bigr)^{d/2} \int\limits_0^t dt_1 \int\limits_0^{t_1} dt_2 \int\limits_0^{t_2} dt_3 \int\limits_0^{t_3} dt_4 \int\!\!\int\!\!\int\!\!\int d\mathbf{x}_1 d\mathbf{x}_2 d\mathbf{x}_3 d\mathbf{x}_4 \\
& \nonumber \;\;\;\;\; \cdot w(t_1,\mathbf{x}_1) w(t_2,\mathbf{x}_2) w(t_3,\mathbf{x}_3) w(t_4,\mathbf{x}_4) \cdot e^{- i (\mathbf{p}_f - \mathbf{q}_1) \mathbf{x}_1 } e^{- i (\mathbf{q}_1 - \mathbf{q}_2) \mathbf{x}_2 } e^{- i (\mathbf{q}_2 - \mathbf{q}_3) \mathbf{x}_3 } e^{- i (\mathbf{q}_3 - \mathbf{p}_i) \mathbf{x}_4 } e^{-\frac{\mathbf{x}_4^2}{2\alpha}} \Bigl]\delta_{\mu\nu}.
\end{align}
\end{subequations}
\end{widetext}
The next step is to compute the transition probability which consists of three terms
\begin{align}
 & P_{M_{\mu} \rightarrow M_{\nu}}(t) = P^{(0)}_{M_{\mu} \rightarrow M_{\nu}}(t) + P^{(1)}_{M_{\mu} \rightarrow M_{\nu}}(t) \\
 & \nonumber \;\;\;\;\; + P^{(2)}_{M_{\mu} \rightarrow M_{\nu}}(t),
\end{align}
where each term corresponds to zeroth, first and second order by time
\begin{widetext}
\begin{subequations}
\begin{align}
 P^{(0)}_{M_{\mu} \rightarrow M_{\nu}}(\mathbf{p}_i, \alpha ; t) & = \frac{1}{(2\pi)^d} \int d\mathbf{p}_f \mathbb{E} \Bigl[T_{\mu\nu}^{(0)}(\mathbf{p}_f, \mathbf{p}_i, \alpha ; t) T_{\mu\nu}^{(0)*}(\mathbf{p}_f, \mathbf{p}_i, \alpha ; t)\Bigr], \\
 P^{(1)}_{M_{\mu} \rightarrow M_{\nu}}(\mathbf{p}_i, \alpha ; t) & = \frac{1}{(2\pi)^d} \int d\mathbf{p}_f \mathbb{E} \Bigl[T_{\mu\nu}^{(0)}(\mathbf{p}_f, \mathbf{p}_i, \alpha ; t) T_{\mu\nu}^{(2)*}(\mathbf{p}_f, \mathbf{p}_i, \alpha ; t) + T_{\mu\nu}^{(2)}(\mathbf{p}_f, \mathbf{p}_i, \alpha ; t) T_{\mu\nu}^{(0)*}(\mathbf{p}_f, \mathbf{p}_i, \alpha ; t) \\
 & \nonumber \;\;\;\;\; + T_{\mu\nu}^{(1)}(\mathbf{p}_f, \mathbf{p}_i, \alpha ; t) T_{\mu\nu}^{(1)*}(\mathbf{p}_f, \mathbf{p}_i, \alpha ; t)\Bigr], \\
 P^{(2)}_{M_{\mu} \rightarrow M_{\nu}}(\mathbf{p}_i, \alpha ; t) & = \frac{1}{(2\pi)^d} \int d\mathbf{p}_f \mathbb{E} \Bigl[T_{\mu\nu}^{(0)}(\mathbf{p}_f, \mathbf{p}_i, \alpha ; t) T_{\mu\nu}^{(4)*}(\mathbf{p}_f, \mathbf{p}_i, \alpha ; t) + T_{\mu\nu}^{(4)}(\mathbf{p}_f, \mathbf{p}_i, \alpha ; t) T_{\mu\nu}^{(0)*}(\mathbf{p}_f, \mathbf{p}_i, \alpha ; t) \\
 & \nonumber \;\;\;\;\; + T_{\mu\nu}^{(1)}(\mathbf{p}_f, \mathbf{p}_i, \alpha ; t) T_{\mu\nu}^{(3)*}(\mathbf{p}_f, \mathbf{p}_i, \alpha ; t) + T_{\mu\nu}^{(3)}(\mathbf{p}_f, \mathbf{p}_i, \alpha ; t) T_{\mu\nu}^{(1)*}(\mathbf{p}_f, \mathbf{p}_i, \alpha ; t) \\
 & \nonumber \;\;\;\;\; + T_{\mu\nu}^{(2)}(\mathbf{p}_f, \mathbf{p}_i, \alpha ; t) T_{\mu\nu}^{(2)*}(\mathbf{p}_f, \mathbf{p}_i, \alpha ; t)\Bigr].
\end{align}
\end{subequations}
\end{widetext}
First term is trivial and given by:
\begin{align}
& P^{(0)}_{M_{\mu}\rightarrow M_{\nu}}(t) \\
& \nonumber = \Bigl(\frac{2\sqrt{\alpha\pi}}{2\pi} \Bigr)^d \int d\mathbf{p}_f  e^{-\alpha(\mathbf{p}_f - \mathbf{p}_i)^2} \delta_{\mu\nu} = \delta_{\mu\nu}.
\end{align}
Second term consists of three components, $T^{(0)*}_{\mu\nu} T^{(2)}_{\mu\nu}$, $T^{(2)*}_{\mu\nu} T^{(0)}_{\mu\nu}$ and $T^{(1)*}_{\mu\nu} T^{(1)}_{\mu\nu}$, where the first two components result in
\begin{widetext}
\begin{align}
& \nonumber \frac{1}{(2\pi)^d} \int d\mathbf{p}_f \mathbb{E} \Bigl[ T^{(0)*}_{\mu\nu}(\mathbf{p}_f, \mathbf{p}_i, \alpha ; t) T^{(2)}_{\mu\nu}(\mathbf{p}_f, \mathbf{p}_i, \alpha ; t) + T^{(2)*}_{\mu\nu}(\mathbf{p}_f, \mathbf{p}_i, \alpha ; t) T^{(0)}_{\mu\nu}(\mathbf{p}_f, \mathbf{p}_i, \alpha ; t) \Bigr] \\
& \nonumber = -2\delta_{\mu\nu} \frac{\gamma m_\mu^2}{m_0^2} \Bigl( \frac{\sqrt{2} (\alpha\pi)^{1/4}}{(2\pi)^2 (\alpha\pi)^{1/4}} \Bigr)^d \int\!\!\int d\mathbf{p}_f d\mathbf{q} \int\!\!\int d\mathbf{x}_1 d\mathbf{x}_2 \; \cos \Bigl[ (\mathbf{p}_f - \mathbf{q}) \mathbf{x}_1 + (\mathbf{q} - \mathbf{p}_i) \mathbf{x}_2 \Bigr] e^{-\frac{\mathbf{x}_2^2}{2\alpha}} \\
& \nonumber \;\;\;\;\; \cdot e^{-\frac{\alpha}{2}(\mathbf{p}_f - \mathbf{p}_i)^2} \int\limits_0^t dt_1 \int\limits_0^{t_1} dt_2 \mathbb{E} [w(t_1,\mathbf{x}_1) w(t_2,\mathbf{x}_2)]  \\
& \nonumber = -\delta_{\mu\nu} \gamma  \frac{m_\mu^2}{m_0^2} \frac{1}{(\sqrt{4\pi} r_C)^d} \Bigl(\frac{\sqrt{2}}{(2\pi)^{2}}\Bigr)^d \int\!\!\int d\mathbf{p}_f d\mathbf{q} \int\!\!\int d\mathbf{x}_1 d\mathbf{x}_2 \; e^{-\frac{(\mathbf{x}_1-\mathbf{x}_2)^2}{4r_C^2}} \Bigl[ e^{i (\mathbf{p}_f - \mathbf{q}) \mathbf{x}_1 } e^{i (\mathbf{q} - \mathbf{p}_i) \mathbf{x}_2 } \\
& \nonumber \;\;\;\;\; + e^{- i (\mathbf{p}_f - \mathbf{q}) \mathbf{x}_1 } e^{- i (\mathbf{q} - \mathbf{p}_i) \mathbf{x}_2 } \Bigr] e^{-\frac{\mathbf{x}_2^2}{2\alpha}} e^{-\frac{\alpha}{2}(\mathbf{p}_f - \mathbf{p}_i)^2} \cdot C_{2,0}^{(1)} (t) \\
& \nonumber = -2 \delta_{\mu\nu} \frac{1}{(\sqrt{4\pi}r_C)^d} \frac{\gamma m_\mu^2}{m_0^2} \Bigl( 1 - \theta(0) \Bigr) \cdot t.
\end{align}
\end{widetext}
The third component equals to
\begin{widetext}
\begin{align}
 & \nonumber \frac{1}{(2\pi)^d} \int d\mathbf{p}_f \mathbb{E} \Bigl[ T^{(1)*}_{\mu\nu}(\mathbf{p}_f, \mathbf{p}_i, \alpha ; t) T^{(1)}_{\mu\nu}(\mathbf{p}_f, \mathbf{p}_i, \alpha ; t) \Bigr] \\
 & \nonumber = \delta_{\mu\nu} \frac{\gamma m_\mu^2}{m_0^2} \Bigl(\frac{1}{2\pi\sqrt{\alpha\pi}}\Bigr)^d\int d\mathbf{p}_f \int\!\!\int d\mathbf{x}_1 d\mathbf{x}_2 \; e^{- i (\mathbf{p}_f - \mathbf{p}_i) (\mathbf{x}_1 - \mathbf{x}_2) } e^{-\frac{\mathbf{x}_1^2+\mathbf{x}_2^2}{2\alpha}} \int\limits_0^t dt_1 \int\limits_0^t dt_2 \mathbb{E} [w(t_1,\mathbf{x}_1) w(t_2,\mathbf{x}_2)] \\
& \nonumber = \delta_{\mu\nu} \frac{\gamma m_\mu^2}{m_0^2} \frac{1}{(\sqrt{4\pi} r_C)^d} \Bigl(\frac{1}{2\pi\sqrt{\alpha\pi}}\Bigr)^d \int d\mathbf{p}_f \int\!\!\int d\mathbf{x}_1 d\mathbf{x}_2 \; e^{-\frac{(\mathbf{x}_1-\mathbf{x}_2)^2}{4r_C^2}} e^{- i (\mathbf{p}_f - \mathbf{p}_i) (\mathbf{x}_1 - \mathbf{x}_2) } e^{-\frac{\mathbf{x}_1^2+\mathbf{x}_2^2}{2\alpha}} \cdot C_{1,1}^{(1)} (t) \\
& \nonumber = \delta_{\mu\nu} \frac{1}{(\sqrt{4\pi}r_C)^d} \frac{\gamma m_\mu^2}{m_0^2} \cdot t.
\end{align}
\end{widetext}
Consequently:
\begin{align}
& P^{(1)}_{M_\mu\rightarrow M_\nu}(t) = -\delta_{\mu\nu} \frac{1}{(\sqrt{4\pi}r_C)^d}\frac{\gamma m_\mu^2}{m_0^2} \Bigl(1 - 2\theta(0) \Bigr) t .
\end{align}
The computations of the integrals $C_{2,0}^{(1)}(t)$ and $C_{1,1}^{(1)}(t)$ which contain 2-point correlation functions of the noise field, can be found in the~Appendix~\ref{sec:TimeIntegrals}.

Second term consists of five components, $T^{(0)*}T^{(4)}$, $T^{(4)*}T^{(0)}$, $T^{(1)*}T^{(3)}$, $T^{(3)*}T^{(1)}$ and $T^{(2)*}T^{(2)}$, where the first two components result in
\begin{widetext}
\begin{align}
& \nonumber \frac{1}{(2\pi)^d} \int d\mathbf{p}_f \mathbb{E} \Bigl[ T^{(0)*}_{\mu\nu}(\mathbf{p}_f, \mathbf{p}_i, \alpha ; t) T^{(4)}_{\mu\nu}(\mathbf{p}_f, \mathbf{p}_i, \alpha ; t) + T^{(4)*}_{\mu\nu}(\mathbf{p}_f, \mathbf{p}_i, \alpha ; t) T^{(0)}_{\mu\nu}(\mathbf{p}_f, \mathbf{p}_i, \alpha ; t) \Bigr] \\
& \nonumber = 2\delta_{\mu\nu}\frac{\gamma^2 m_\mu^4}{m_0^4} \Bigl( \frac{\sqrt{2}(\alpha\pi)^{1/4}}{(2\pi)^4(\alpha\pi)^{1/4}} \Bigr)^d \int\!\!\int\!\!\int\!\!\int d\mathbf{p}_f d\mathbf{q}_1 d\mathbf{q}_2 d\mathbf{q}_3 \int\!\!\int\!\!\int\!\!\int d\mathbf{x}_1 d\mathbf{x}_2 d\mathbf{x}_3 d\mathbf{x}_4 \; \cos \Bigl[ (\mathbf{p}_f - \mathbf{q}_1) \mathbf{x}_1 + (\mathbf{q}_1 - \mathbf{q}_2) \mathbf{x}_2 \\
& \nonumber \;\;\;\;\; + (\mathbf{q}_2 - \mathbf{q}_3) \mathbf{x}_3 + (\mathbf{q}_3 - \mathbf{p}_i) \mathbf{x}_4 \Bigr] e^{-\frac{\mathbf{x}_4^2}{2\alpha}} e^{-\frac{\alpha}{2}(\mathbf{p}_f - \mathbf{p}_i)^2} \int\limits_0^t dt_1 \int\limits_0^{t_1} dt_2 \int\limits_0^{t_2} dt_3 \int\limits_0^{t_3} dt_4 \mathbb{E} [w(t_1, \mathbf{x}_1) w(t_2, \mathbf{x}_2) w(t_3, \mathbf{x}_3) w(t_4, \mathbf{x}_4)] \\
& \nonumber = \delta_{\mu\nu} \frac{1}{(4\pi r_C^2)^d}\frac{\gamma^2 m_\mu^4}{m_0^4} \Bigl(\frac{\sqrt{2}}{(2\pi)^4}\Bigr)^d\int\!\!\int\!\!\int\!\!\int d\mathbf{p}_f d\mathbf{q}_1 d\mathbf{q}_2 d\mathbf{q}_3 \int\!\!\int\!\!\int\!\!\int d\mathbf{x}_1 d\mathbf{x}_2 d\mathbf{x}_3 d\mathbf{x}_4 \\
& \nonumber \;\;\;\;\; \cdot \Bigl[ e^{i (\mathbf{p}_f - \mathbf{q}_1) \mathbf{x}_1 } e^{i (\mathbf{q}_1 - \mathbf{q}_2) \mathbf{x}_2 } e^{i (\mathbf{q}_2 - \mathbf{q}_3) \mathbf{x}_3 } e^{i (\mathbf{q}_3 - \mathbf{p}_i) \mathbf{x}_4 } + e^{- i (\mathbf{p}_f - \mathbf{q}_1) \mathbf{x}_1 } e^{- i (\mathbf{q}_1 - \mathbf{q}_2) \mathbf{x}_2 } e^{- i (\mathbf{q}_2 - \mathbf{q}_3) \mathbf{x}_3 } e^{- i (\mathbf{q}_3 - \mathbf{p}_i) \mathbf{x}_4 } \Bigr] e^{-\frac{\mathbf{x}_4^2}{2\alpha}} e^{-\frac{\alpha}{2}(\mathbf{p}_f - \mathbf{p}_i)^2} \\
& \nonumber \;\;\;\;\; \cdot \Bigl[ e^{-\frac{(\mathbf{x}_1-\mathbf{x}_2)^2}{4r_C^2}} e^{-\frac{(\mathbf{x}_3-\mathbf{x}_4)^2}{4r_C^2}} U^{4,0}_1(t) + e^{-\frac{(\mathbf{x}_1-\mathbf{x}_3)^2}{4r_C^2}} e^{-\frac{(\mathbf{x}_2-\mathbf{x}_4)^2}{4r_C^2}} U^{4,0}_2(t) + e^{-\frac{(\mathbf{x}_1-\mathbf{x}_4)^2}{4r_C^2}} e^{-\frac{(\mathbf{x}_2-\mathbf{x}_3)^2}{4r_C^2}} U^{4,0}_3(t) \Bigr] \\
& \nonumber = \delta_{\mu\nu} \frac{1}{(4\pi r_C^2)^d}\frac{\gamma^2 m_\mu^4}{m_0^4} \Bigl(\frac{\sqrt{2}}{(2\pi)^4}\Bigr)^d\int\!\!\int\!\!\int\!\!\int d\mathbf{p}_f d\mathbf{q}_1 d\mathbf{q}_2 d\mathbf{q}_3 \int\!\!\int\!\!\int\!\!\int d\mathbf{x}_1 d\mathbf{x}_2 d\mathbf{x}_3 d\mathbf{x}_4 \\
& \nonumber \;\;\;\;\; \cdot \Bigl[ e^{i (\mathbf{p}_f - \mathbf{q}_1) \mathbf{x}_1 } e^{i (\mathbf{q}_1 - \mathbf{q}_2) \mathbf{x}_2 } e^{i (\mathbf{q}_2 - \mathbf{q}_3) \mathbf{x}_3 } e^{i (\mathbf{q}_3 - \mathbf{p}_i) \mathbf{x}_4 } + e^{- i (\mathbf{p}_f - \mathbf{q}_1) \mathbf{x}_1 } e^{- i (\mathbf{q}_1 - \mathbf{q}_2) \mathbf{x}_2 } e^{- i (\mathbf{q}_2 - \mathbf{q}_3) \mathbf{x}_3 } e^{- i (\mathbf{q}_3 - \mathbf{p}_i) \mathbf{x}_4 } \Bigr] \\
& \nonumber \;\;\;\;\; \cdot e^{-\frac{\mathbf{x}_4^2}{2\alpha}} e^{-\frac{\alpha}{2}(\mathbf{p}_f - \mathbf{p}_i)^2} e^{-\frac{(\mathbf{x}_1-\mathbf{x}_2)^2}{4r_C^2}} e^{-\frac{(\mathbf{x}_3-\mathbf{x}_4)^2}{4r_C^2}} \cdot \frac{1}{2} \Bigl(1 - \theta(0) \Bigr)^2 t^2 \\
& \nonumber = 2\delta_{\mu\nu} \frac{1}{(4\pi r_C^2)^d} \frac{\gamma^2 m_\mu^4}{m_0^4} \cdot \frac{1}{2} \Bigl(1 - \theta(0) \Bigr)^2 \cdot t^2,
\end{align}
\end{widetext}
the second two components result in
\begin{widetext}
\begin{align}
& \nonumber \frac{1}{(2\pi)^d} \int d\mathbf{p}_f \mathbb{E} \Bigl[ T^{(1)*}_{\mu\nu}(\mathbf{p}_f, \mathbf{p}_i, \alpha ; t) T^{(3)}_{\mu\nu}(\mathbf{p}_f, \mathbf{p}_i, \alpha ; t) + T^{(3)*}_{\mu\nu}(\mathbf{p}_f, \mathbf{p}_i, \alpha ; t) T^{(1)}_{\mu\nu}(\mathbf{p}_f, \mathbf{p}_i, \alpha ; t) \Bigr] \\
& \nonumber = -2\delta_{\mu\nu} \frac{\gamma^2 m_\mu^4}{m_0^4} \Bigl(\frac{1}{(2\pi)^3\sqrt{\alpha\pi}}\Bigr)^d \int\!\!\int\!\!\int d\mathbf{p}_f d\mathbf{q}_1 d\mathbf{q}_2 \int\!\!\int\!\!\int\!\!\int d\mathbf{x}_1 d\mathbf{x}_2 d\mathbf{x}_3 d\mathbf{x}_4 \; \cos \Bigl[ (\mathbf{p}_f - \mathbf{q}_1) \mathbf{x}_1 + (\mathbf{q}_1 - \mathbf{q}_2) \mathbf{x}_2 + (\mathbf{q}_2 - \mathbf{p}_i) \mathbf{x}_3  \\
& \nonumber \;\;\;\;\; - (\mathbf{p}_f - \mathbf{p}_i) \mathbf{x}_4 \Bigr] e^{-\frac{\mathbf{x}_3^2 + \mathbf{x}_4^2}{2\alpha}} \int\limits_0^t dt_1 \int\limits_0^{t_1} dt_2 \int\limits_0^{t_2} dt_3 \int\limits_0^t dt_4 \mathbb{E} [w(t_1, \mathbf{x}_1) w(t_2, \mathbf{x}_2) w(t_3, \mathbf{x}_3) w(t_4, \mathbf{x}_4)] \\
& \nonumber = -\delta_{\mu\nu} \frac{1}{(4\pi r_C^2)^d} \frac{\gamma^2 m_\mu^4}{m_0^4} \Bigl(\frac{1}{(2\pi)^3\sqrt{\alpha\pi}}\Bigr)^d \int\!\!\int\!\!\int d\mathbf{p}_f d\mathbf{q}_1 d\mathbf{q}_2 \int\!\!\int\!\!\int\!\!\int d\mathbf{x}_1 d\mathbf{x}_2 d\mathbf{x}_3 d\mathbf{x}_4 \\
& \nonumber \;\;\;\;\; \cdot \Bigl[ e^{i (\mathbf{p}_f - \mathbf{q}_1) \mathbf{x}_1 } e^{i (\mathbf{q}_1 - \mathbf{q}_2) \mathbf{x}_2 } e^{i (\mathbf{q}_2 - \mathbf{p}_i) \mathbf{x}_3 } e^{-i (\mathbf{p}_f - \mathbf{p}_i) \mathbf{x}_4 } + e^{-i (\mathbf{p}_f - \mathbf{q}_1) \mathbf{x}_1 } e^{-i (\mathbf{q}_1 - \mathbf{q}_2) \mathbf{x}_2 } e^{-i (\mathbf{q}_2 - \mathbf{p}_i) \mathbf{x}_3 } e^{i (\mathbf{p}_f - \mathbf{p}_i) \mathbf{x}_4 } \Bigr] e^{-\frac{\mathbf{x}_3^2 + \mathbf{x}_4^2}{2\alpha}} \\
& \nonumber \;\;\;\;\; \cdot \Bigl[ e^{-\frac{(\mathbf{x}_1-\mathbf{x}_2)^2}{4r_C^2}} e^{-\frac{(\mathbf{x}_3-\mathbf{x}_4)^2}{4r_C^2}} U^{3,1}_1(t) + e^{-\frac{(\mathbf{x}_1-\mathbf{x}_3)^2}{4r_C^2}} e^{-\frac{(\mathbf{x}_2-\mathbf{x}_4)^2}{4r_C^2}} U^{3,1}_2(t) + e^{-\frac{(\mathbf{x}_1-\mathbf{x}_4)^2}{4r_C^2}} e^{-\frac{(\mathbf{x}_2-\mathbf{x}_3)^2}{4r_C^2}} U^{3,1}_3(t) \Bigr] \\
& \nonumber = -\delta_{\mu\nu} \frac{1}{(4\pi r_C^2)^d} \frac{\gamma^2 m_\mu^4}{m_0^4} \Bigl(\frac{1}{(2\pi)^3\sqrt{\alpha\pi}}\Bigr)^d \int\!\!\int\!\!\int d\mathbf{p}_f d\mathbf{q}_1 d\mathbf{q}_2 \int\!\!\int\!\!\int\!\!\int d\mathbf{x}_1 d\mathbf{x}_2 d\mathbf{x}_3 d\mathbf{x}_4 \\
& \nonumber \;\;\;\;\; \cdot \Bigl[ e^{i (\mathbf{p}_f - \mathbf{q}_1) \mathbf{x}_1 } e^{i (\mathbf{q}_1 - \mathbf{q}_2) \mathbf{x}_2 } e^{i (\mathbf{q}_2 - \mathbf{p}_i) \mathbf{x}_3 } e^{-i (\mathbf{p}_f - \mathbf{p}_i) \mathbf{x}_4 } + e^{-i (\mathbf{p}_f - \mathbf{q}_1) \mathbf{x}_1 } e^{-i (\mathbf{q}_1 - \mathbf{q}_2) \mathbf{x}_2 } e^{-i (\mathbf{q}_2 - \mathbf{p}_i) \mathbf{x}_3 } e^{i (\mathbf{p}_f - \mathbf{p}_i) \mathbf{x}_4 } \Bigr] e^{-\frac{\mathbf{x}_3^2 + \mathbf{x}_4^2}{2\alpha}} \\
& \nonumber \;\;\;\;\; \cdot \Bigl[ e^{-\frac{(\mathbf{x}_1-\mathbf{x}_2)^2}{4r_C^2}} e^{-\frac{(\mathbf{x}_3-\mathbf{x}_4)^2}{4r_C^2}} + e^{-\frac{(\mathbf{x}_1-\mathbf{x}_4)^2}{4r_C^2}} e^{-\frac{(\mathbf{x}_2-\mathbf{x}_3)^2}{4r_C^2}} \Bigr] \cdot \frac{1}{2} \Bigl(1 - \theta(0) \Bigr) t^2 \\
& \nonumber = -2\delta_{\mu\nu} \frac{1}{(4\pi r_C^2)^d} \frac{\gamma^2 m_\mu^4}{m_0^4} \Bigl(1 - \theta(0) \Bigr) \cdot t^2,
\end{align}
\end{widetext}
and the last component equals to
\begin{widetext}
\begin{align}
& \nonumber \frac{1}{(2\pi)^d} \int d\mathbf{p}_f \mathbb{E} \Bigl[ T^{(2)*}_{\mu\nu}(\mathbf{p}_f, \mathbf{p}_i, \alpha ; t) T^{(2)}_{\mu\nu}(\mathbf{p}_f, \mathbf{p}_i, \alpha ; t) \Bigr] \\
& \nonumber = \delta_{\mu\nu} \frac{\gamma^2 m_\mu^4}{m_0^4} \Bigl(\frac{1}{(2\pi)^3\sqrt{\alpha\pi}}\Bigr)^d \int\!\!\int\!\!\int d\mathbf{p}_f d\mathbf{q}_1 d\mathbf{q}_2 \int\!\!\int\!\!\int\!\!\int d\mathbf{x}_1 d\mathbf{x}_2 d\mathbf{x}_3 d\mathbf{x}_4 \; e^{- i (\mathbf{p}_f - \mathbf{q}_1) \mathbf{x}_1 } e^{- i (\mathbf{q}_1 - \mathbf{p}_i) \mathbf{x}_2 } e^{-\frac{\mathbf{x}_2^2}{2\alpha}}\\
& \nonumber \;\;\;\;\; \cdot e^{i (\mathbf{p}_f - \mathbf{q}_2) \mathbf{x}_3 } e^{i (\mathbf{q}_2 - \mathbf{p}_i) \mathbf{x}_4 } e^{-\frac{\mathbf{x}_4^2}{2\alpha}} \int\limits_0^t dt_1 \int\limits_0^{t_1} dt_2 \int\limits_0^{t} dt_3 \int\limits_0^{t_3} dt_4 \mathbb{E} [w(t_1, \mathbf{x}_1) w(t_2, \mathbf{x}_2) w(t_3, \mathbf{x}_3) w(t_4, \mathbf{x}_4)] \\
& \nonumber = \delta_{\mu\nu} \frac{1}{(4\pi r_C^2)^d} \frac{\gamma^2 m_\mu^4}{m_0^4} \Bigl(\frac{1}{(2\pi)^3\sqrt{\alpha\pi}}\Bigr)^d \int\!\!\int\!\!\int d\mathbf{p}_f d\mathbf{q}_1 d\mathbf{q}_2 \int\!\!\int\!\!\int\!\!\int d\mathbf{x}_1 d\mathbf{x}_2 d\mathbf{x}_3 d\mathbf{x}_4 \; e^{- i (\mathbf{p}_f - \mathbf{q}_1) \mathbf{x}_1 } e^{- i (\mathbf{q}_1 - \mathbf{p}_i) \mathbf{x}_2 } e^{-\frac{\mathbf{x}_2^2}{2\alpha}}\\
& \nonumber \;\;\;\;\; \cdot e^{- i (\mathbf{p}_f - \mathbf{q}_1) \mathbf{x}_1 } e^{- i (\mathbf{q}_1 - \mathbf{p}_i) \mathbf{x}_2 } e^{-\frac{\mathbf{x}_2^2}{2\alpha}} e^{i (\mathbf{p}_f - \mathbf{q}_2) \mathbf{x}_3 } e^{i (\mathbf{q}_2 - \mathbf{p}_i) \mathbf{x}_4 } e^{-\frac{\mathbf{x}_4^2}{2\alpha}} \\
& \nonumber \;\;\;\;\; \cdot \Bigl[ e^{-\frac{(\mathbf{x}_1-\mathbf{x}_2)^2}{4r_C^2}} e^{-\frac{(\mathbf{x}_3-\mathbf{x}_4)^2}{4r_C^2}} U^{2,2}_1(t) + e^{-\frac{(\mathbf{x}_1-\mathbf{x}_3)^2}{4r_C^2}} e^{-\frac{(\mathbf{x}_2-\mathbf{x}_4)^2}{4r_C^2}} U^{2,2}_2(t) + e^{-\frac{(\mathbf{x}_1-\mathbf{x}_4)^2}{4r_C^2}} e^{-\frac{(\mathbf{x}_2-\mathbf{x}_3)^2}{4r_C^2}} U^{2,2}_3(t) \Bigr] \\
& \nonumber = \delta_{\mu\nu} \frac{1}{(4\pi r_C^2)^d} \frac{\gamma^2 m_\mu^4}{m_0^4} \Bigl(\frac{1}{(2\pi)^3\sqrt{\alpha\pi}}\Bigr)^d \int\!\!\int\!\!\int d\mathbf{p}_f d\mathbf{q}_1 d\mathbf{q}_2 \int\!\!\int\!\!\int\!\!\int d\mathbf{x}_1 d\mathbf{x}_2 d\mathbf{x}_3 d\mathbf{x}_4 \; e^{- i (\mathbf{p}_f - \mathbf{q}_1) \mathbf{x}_1 } e^{- i (\mathbf{q}_1 - \mathbf{p}_i) \mathbf{x}_2 } e^{-\frac{\mathbf{x}_2^2}{2\alpha}}\\
& \nonumber \;\;\;\;\; \cdot e^{- i (\mathbf{p}_f - \mathbf{q}_1) \mathbf{x}_1 } e^{- i (\mathbf{q}_1 - \mathbf{p}_i) \mathbf{x}_2 } e^{-\frac{\mathbf{x}_2^2}{2\alpha}} e^{i (\mathbf{p}_f - \mathbf{q}_2) \mathbf{x}_3 } e^{i (\mathbf{q}_2 - \mathbf{p}_i) \mathbf{x}_4 } e^{-\frac{\mathbf{x}_4^2}{2\alpha}} \\
& \nonumber \;\;\;\;\; \Bigl[ e^{-\frac{(\mathbf{x}_1-\mathbf{x}_2)^2}{4r_C^2}} e^{-\frac{(\mathbf{x}_3-\mathbf{x}_4)^2}{4r_C^2}} \cdot \Bigl(1 - \theta(0) \Bigr)^2 + e^{-\frac{(\mathbf{x}_1-\mathbf{x}_3)^2}{4r_C^2}} e^{-\frac{(\mathbf{x}_2-\mathbf{x}_4)^2}{4r_C^2}} \cdot \frac{1}{2} \Bigr] t^2 \\
& \nonumber = \delta_{\mu\nu} \frac{1}{(4\pi r_C^2)^d} \frac{\gamma^2 m_\mu^4}{m_0^4} \Bigl( \Bigl(1 - \theta(0) \Bigr)^2 + \frac{1}{2} \Bigr) \cdot t^2.
\end{align}
\end{widetext}
where
\begin{align}
& \nonumber U^{4,0}_1(t) + U^{4,0}_2(t) + U^{4,0}_3(t) \equiv C^{(2)}_{4,0}(t), \\
& \nonumber U^{3,1}_1(t) + U^{3,1}_2(t) + U^{3,1}_3(t) \equiv C^{(2)}_{3,1}(t), \\
& \nonumber U^{2,2}_1(t) + U^{2,2}_2(t) + U^{2,2}_3(t) \equiv C^{(2)}_{2,2}(t)
\end{align}
correspond to the integrals of the 4-point correlation functions of the noise field, which are calculated in Appendix~\ref{sec:TimeIntegrals}.

Consequently, the component of the transition probabilities, which corresponds to the second order by time $t$, equals to
\begin{align}\label{SecondComponentProbCSL}
& P^{(2)}_{M_\mu\rightarrow M_\nu}(t) \\
& \nonumber = \delta_{\mu\nu}\frac{1}{(4\pi r_C^2)^d} \frac{\gamma^2 m_\mu^4}{m_0^4} \Bigl(2\theta(0) (\theta(0) - 1 ) + \frac{1}{2}\Bigr) t^2 \\
& \nonumber = \delta_{\mu\nu}\frac{1}{2} \frac{1}{(4\pi r_C^2)^d} \frac{\gamma^2 m_\mu^4}{m_0^4} \Bigl(1 - 2\theta(0)\Bigr)^2 t^2 .
\end{align}
Finally, collecting all the calculated terms~(\ref{ZerothComponentProbCSL})--(\ref{SecondComponentProbCSL}), we obtain the transition probabilities for mass eigenstates
\begin{align}
& P_{M_\mu\rightarrow M_\nu} (t) = \Bigl[ 1 - \gamma \frac{m_\mu^2}{m_0^2} \frac{1}{(\sqrt{4\pi}r_C)^d} \Bigl(1 - 2\theta(0) \Bigr) t \\
& \nonumber \;\;\;\;\; + \frac{\gamma^2}{2} \frac{m_\mu^4}{m_0^4} \frac{1}{(4\pi r_C^2)^d} \Bigl(1 - 2\theta(0)\Bigr)^2 t^2 \Bigr] \delta_{\mu\nu}.
\end{align}

\subsection{Transition probabilities for the flavor states}
We perform the computations in the same manner as was done in~Appendix~\ref{sec:QMUPL_Computations} for the QMUPL model, and expand the probabilities for the flavor states for the mass-proportional CSL model in the following form
\begin{align}
& P_{M^0 \rightarrow M^0/\bar{M}^0}(t) = \sum_{\mu, \mu'} \alpha_{\mu} \beta^*_{\mu} \alpha^*_{\mu'} \beta_{\mu'} P_{\mu \mu'}(t) \\
& \nonumber = \frac{1}{4} (P_{HH}(t) \pm P_{HL}(t) \pm P_{LH}(t) + P_{LL}(t)),
\end{align}
where terms with same indexes $P_{aa}$ and different ones $P_{ab}$ are equal to
\begin{align}
& P_{aa}(t) = 1 - \frac{1}{(\sqrt{4\pi}r_C)^d} \frac{\gamma m_a^2}{m_0^2} \Bigl( 1 - 2\theta(0)\Bigr) t \\
& \nonumber \;\;\;\;\; + \frac{1}{2} \frac{1}{(4\pi r_C^2)^d} \frac{\gamma^2 m_a^4}{m_0^4} \Bigl( 1 - 2\theta(0)\Bigr)^2 t^2, \\
& P_{ab}(t) = e^{-i(m_a - m_b)t} \\
& \nonumber \;\;\;\;\; \cdot \Biggl\{ 1 - \frac{1}{(\sqrt{4\pi}r_C)^d} \frac{\gamma}{m_0^2} \Bigl( (m_a^2 + m_b^2)\Bigl( 1 - \theta(0) \Bigr) - m_a m_b\Bigr) t \\
& \nonumber \;\;\;\;\; + \frac{1}{(4\pi r_C^2)^d} \frac{\gamma^2}{m_0^4} \Bigl[(m_a^4 + m_b^4) \Bigl( 1 - \theta(0) \Bigr)^2 \\
& \nonumber \;\;\;\;\; - 2(m_a^3 m_b + m_a m_b^3) \Bigl( 1 - \theta(0) \Bigr) \\
& \nonumber \;\;\;\;\; + 2m_a^2 m_b^2 \Bigl(\Bigl( 1 - \theta(0) \Bigr)^2 + \frac{1}{2}\Bigr) \Bigr] t^2 \Biggr\}.
\end{align}
Putting the terms together, we finally obtain the transition probability for the flavor states for the mass-proportional CSL model
\begin{widetext}
\begin{align}
 & P_{M^0 \rightarrow M^0/\bar{M}^0}(t) = \frac{1}{2} \Biggl\{ 1 - \frac{1}{2} \frac{1}{(\sqrt{4\pi}r_C)^d} \frac{\gamma (m_H^2+m_L^2)}{m_0^2} \Bigl( 1 - 2\theta(0) \Bigr) \cdot t \\
 & \nonumber \;\;\;\;\; + \frac{1}{4} \frac{1}{(4\pi r_C^2)^d} \frac{\gamma^2 (m_H^4 + m_L^4)}{m_0^4} \Bigl( 1 - 2\theta(0) \Bigr)^2 \cdot t^2 \pm \Biggl[1 - \frac{1}{(\sqrt{4\pi}r_C)^d} \frac{\gamma}{m_0^2} \Bigl( (m_H^2 + m_L^2)\Bigl( 1 - \theta(0) \Bigr) - m_H m_L\Bigr) \cdot t \\
 & \nonumber \;\;\;\;\; + \frac{1}{2} \frac{1}{(4\pi r_C^2)^d} \frac{\gamma^2}{m_0^4} \Bigl( (m_H^4 + m_L^4)\Bigl( 1 - \theta(0) \Bigr)^2 - 2m_H m_L (m_H^2 + m_L^2)\Bigl( 1 - \theta(0) \Bigr) + 2m_H^2 m_L^2 \Bigl(\Bigl( 1 - \theta(0) \Bigr)^2 + \frac{1}{2} \Bigr) \Bigr) \cdot t^2 \Biggr] \\
 & \nonumber \;\;\;\;\; \cdot\cos\Bigl[(m_H - m_L)t\Bigr]\Biggr\}.
\end{align}
\end{widetext}
Taking decay into account
\begin{widetext}
\begin{align}
 & P_{M^0 \rightarrow M^0/\bar{M}^0}(t) = \frac{1}{4} \Biggl\{ e^{-\Gamma_H t}+e^{-\Gamma_L t} - \frac{1}{(\sqrt{4\pi}r_C)^d} \frac{\gamma}{m_0^2} (m_H^2 e^{-\Gamma_H t}+ m_L^2 e^{-\Gamma_L t})\Bigl(1 - 2\theta(0)\Bigr) \cdot t \\
 & \nonumber \;\;\;\;\; + \frac{1}{2} \frac{1}{(4\pi r_C^2)^d} \frac{\gamma^2}{m_0^4} (m_H^4 e^{-\Gamma_H t}+ m_L^4e^{-\Gamma_L t}) \Bigl( 1 - 2\theta(0) \Bigr)^2 \cdot t^2 \\
 & \nonumber \;\;\;\;\; \pm 2\Biggl[1 - \frac{1}{(\sqrt{4\pi}r_C)^d}\frac{\gamma}{m_0^2} \Bigl( (m_H^2 + m_L^2)\Bigl( 1 - \theta(0) \Bigr) - m_H m_L\Bigr) \cdot t + \frac{1}{2} \frac{1}{(4\pi r_C^2)^d} \frac{\gamma^2}{m_0^4} \Bigl( (m_H^4 + m_L^4)\Bigl( 1 - \theta(0) \Bigr)^2 \\
 & \nonumber \;\;\;\;\; - 2m_H m_L (m_H^2 + m_L^2)\Bigl( 1 - \theta(0) \Bigr) + 2m_H^2 m_L^2 \Bigl(\Bigl( 1 - \theta(0) \Bigr)^2 + \frac{1}{2} \Bigr) \Bigr) \cdot t^2 \Biggr]\\
 & \nonumber \;\;\;\;\; \cdot\cos\Bigl[(m_H - m_L)t\Bigr]\cdot e^{-\frac{\Gamma_H+\Gamma_L}{2} t}\Biggr\}.
\end{align}
\end{widetext}

\section{Correlation functions of the noise field}\label{sec:TimeIntegrals}
\subsection{Calculations with a 2-point correlation function}
First-order components of the transition probabilities contain a $2$-point correlation function of the noise. In the computations for the QMUPL model the noise is assumed to be a white one, i.e. any random process is uncorrelated to the random process at a later time point. Mathematically, one defines $\mathbb{E}[w(t_1)w(t_2)] = \frac{1}{2\pi} \int_{-\infty}^{\infty}d\omega\; e^{i \omega (t_1-t_2)}= \delta(t_1-t_2)$. In our computations two different integrals have to be computed (corresponding to $T^{(0)*}T^{(2)}$ and $T^{(2)*}T^{(0)}$, respectively):
\begin{align}
 & C^{(1)}_{2,0}(t) = \int\limits_0^t dt_1 \int\limits_0^{t_1} dt_2\, \delta(t_1 - t_2) \\
 & \nonumber = \int\limits_0^t dt_1 ( \theta(t_1) - \theta(0)) = (1-\theta(0)) t,
\end{align}
and the second one corresponds to the component $T^{(1)*}T^{(1)}$:
\begin{align}
 & C^{(1)}_{1,1}(t) = \int\limits_0^t dt_1 \int\limits_0^t dt_2 \delta(t_1 - t_2) \\
 & \nonumber = \int\limits_0^t dt_1 ( \theta(t_1) - \theta(t_1-t)) = t.
\end{align}
In the $d$-dimensional case we define $\mathbb{E}[\mathbf{w}(t_1) \cdot \mathbf{w} (t_2)] = \delta(t_1-t_2)$.

\subsection{Calculations with a 4-point correlation function}
Second-order components of the transition probabilities contain integrals of a 4-point correlation function of the noise field
\begin{align}
 & \nonumber C^{(2)}_{4,0}(t) = \int\limits_0^t dt_1 \int\limits_0^{t_1} dt_2 \int\limits_0^{t_2} dt_3 \int\limits_0^{t_3} dt_4 \; \mathbb{E}[w(t_1)w(t_2)w(t_3)w(t_4)], \\
 & \nonumber C^{(2)}_{3,1}(t) = \int\limits_0^t dt_1 \int\limits_0^{t_1} dt_2 \int\limits_0^{t_2} dt_3 \int\limits_0^{t} dt_4 \; \mathbb{E}[w(t_1)w(t_2)w(t_3)w(t_4)], \\
 & \nonumber C^{(2)}_{2,2}(t) = \int\limits_0^t dt_1 \int\limits_0^{t_1} dt_2 \int\limits_0^{t} dt_3 \int\limits_0^{t_3} dt_4 \; \mathbb{E}[w(t_1)w(t_2)w(t_3)w(t_4)].
\end{align}
Since the noise field is assumed to be a Gaussian white noise field, its 4th cumulant is equal to zero, $\kappa ( w(t_1) w(t_2) w(t_3) w(t_4) ) = 0$. On the other hand, odd moments of the Gaussian noise are equal to zero as well, therefore it is possible to reformulate its 4-point correlation function as a combination of 2-point correlation functions:
\begin{align}
& \nonumber \mathbb{E}[w(t_1)w(t_2)w(t_3)w(t_4)] = \mathbb{E}[w(t_1)w(t_2)]\mathbb{E}[w(t_3)w(t_4)] \\
& \nonumber \;\;\;\;\; + \mathbb{E}[w(t_1)w(t_3)]\mathbb{E}[w(t_2)w(t_4)] \\
& \nonumber \;\;\;\;\; + \mathbb{E}[w(t_1)w(t_4)]\mathbb{E}[w(t_2)w(t_3)].
\end{align}
Accordingly, each second-order components of the transition probability contains three integrals of two 2-point correlation functions
\begin{align}
& \nonumber C^{(2)}_{4,0}(t) \equiv U^{4,0}_1(t) + U^{4,0}_2(t) + U^{4,0}_3(t), \\
& \nonumber C^{(2)}_{3,1}(t) \equiv U^{3,1}_1(t) + U^{3,1}_2(t) + U^{3,1}_3(t), \\
& \nonumber C^{(2)}_{2,2}(t) \equiv U^{2,2}_1(t) + U^{2,2}_2(t) + U^{2,2}_3(t).
\end{align}
For the components $T^{(0)*}T^{(4)}$ and $T^{(4)*}T^{(0)}$ the first integral is equal to:
\begin{align}
& \nonumber U^{4,0}_1(t) = \int\limits_0^t dt_1 \int\limits_0^{t_1} dt_2 \int\limits_0^{t_2} dt_3 \int\limits_0^{t_3} dt_4 \delta(t_1 - t_2) \delta(t_3 - t_4) \\
& \nonumber = \int\limits_0^t dt_1 \int\limits_0^{t_1} dt_2 \int\limits_0^{t_2} dt_3 (\theta(t_3) - \theta(0)) \delta(t_1 - t_2) \\
& \nonumber = \int\limits_0^t dt_1 \int\limits_0^{t_1} dt_2 \; t_2 (\theta(t_2) - \theta(0)) \delta(t_1 - t_2) \\
& \nonumber = \int\limits_0^t dt_1 \; t_1 (\theta(t_1) - \theta(0))^2 = \frac{1}{2} (1 - \theta(0))^2 t^2.
\end{align}
Second integral:
\begin{align}
& \nonumber U^{4,0}_2(t)  = \int\limits_0^t dt_1 \int\limits_0^{t_1} dt_2 \int\limits_0^{t_2} dt_3 \int\limits_0^{t_3} dt_4 \delta(t_1 - t_3) \delta(t_2 - t_4) \\
& \nonumber = \int\limits_0^t dt_1 \int\limits_0^{t_1} dt_2 \int\limits_0^{t_2} dt_3 (\theta(t_2) - \theta(t_2 - t_3)) \delta(t_1 - t_3)  \\
& \nonumber = \int\limits_0^t dt_1 \int\limits_0^{t_1} dt_2 (\theta(t_2) - \theta(t_2 - t_1)) (\theta(t_1) - \theta(t_1 - t_2)) \\
& \nonumber = \int\limits_0^t dt_1 \; t_1 (\theta^2(t_1) - \theta(t_1)) = 0.
\end{align}
Third integral:
\begin{align}
& \nonumber U^{4,0}_3(t) = \int\limits_0^t dt_1 \int\limits_0^{t_1} dt_2 \int\limits_0^{t_2} dt_3 \int\limits_0^{t_3} dt_4 \delta(t_1 - t_4) \delta(t_2 - t_3) \\
& \nonumber = \int\limits_0^t dt_1 \int\limits_0^{t_1} dt_2 \int\limits_0^{t_2} dt_3 (\theta(t_1) - \theta(t_1 - t_3)) \delta(t_2 - t_3) \\
& \nonumber = \int\limits_0^t dt_1 \int\limits_0^{t_1} dt_2 (\theta(t_1) - \theta(t_1 - t_2)) (\theta(t_2) - \theta(0)) \\
& \nonumber = \int\limits_0^t dt_1 \; t_1 (\theta^2(t_1) - \theta(t_1)) = 0.
\end{align}
For the components $T^{(1)*}T^{(3)}$ and $T^{(3)*}T^{(1)}$ the first integral is equal to:
\begin{align}
& \nonumber U^{3,1}_1(t) = \int\limits_0^t dt_1 \int\limits_0^{t_1} dt_2 \int\limits_0^{t_2} dt_3 \int\limits_0^{t} dt_4 \delta(t_1 - t_2) \delta(t_3 - t_4) \\
& \nonumber = \int\limits_0^t dt_1 \int\limits_0^{t_1} dt_2 \int\limits_0^{t_2} dt_3 (\theta(t_3) - \theta(t_3-t)) \delta(t_1 - t_2) \\
& \nonumber = \int\limits_0^t dt_1 \int\limits_0^{t_1} dt_2 \; (t_2 \theta(t_2) - (t_2 - t)\theta(t_2-t)) \delta(t_1 - t_2) \\
& \nonumber = \int\limits_0^t dt_1 \; (t_1 \theta(t_1) - (t_1 - t)\theta(t_1-t))(\theta(t_1) - \theta(0))\\
& \nonumber = \frac{1}{2} (1-\theta(0)) t^2.
\end{align}
Second integral:
\begin{align}
& \nonumber U^{3,1}_2(t) = \int\limits_0^t dt_1 \int\limits_0^{t_1} dt_2 \int\limits_0^{t_2} dt_3 \int\limits_0^{t} dt_4 \delta(t_1 - t_3) \delta(t_2 - t_4) \\
& \nonumber = \int\limits_0^t dt_1 \int\limits_0^{t_1} dt_2 \int\limits_0^{t_2} dt_3 (\theta(t_2) - \theta(t_2 - t)) \delta(t_1 - t_3)  \\
& \nonumber = \int\limits_0^t dt_1 \int\limits_0^{t_1} dt_2 (\theta(t_2) - \theta(t_2 - t)) (\theta(t_1) - \theta(t_1 - t_2)) \\
& \nonumber = \int\limits_0^t dt_1 \; t_1 (\theta^2(t_1) - \theta(t_1)) = 0.
\end{align}
Third integral:
\begin{align}
& \nonumber U^{3,1}_3(t) = \int\limits_0^t dt_1 \int\limits_0^{t_1} dt_2 \int\limits_0^{t_2} dt_3 \int\limits_0^{t} dt_4 \delta(t_1 - t_4) \delta(t_2 - t_3) \\
& \nonumber = \int\limits_0^t dt_1 \int\limits_0^{t_1} dt_2 \int\limits_0^{t_2} dt_3 (\theta(t_1) - \theta(t_1 - t)) \delta(t_2 - t_3) \\
& \nonumber = \int\limits_0^t dt_1 \int\limits_0^{t_1} dt_2 (\theta(t_1) - \theta(t_1 - t)) (\theta(t_2) - \theta(0)) \\
& \nonumber = \int\limits_0^t dt_1 \; t_1 (\theta(t_1) - \theta(t_1 - t))(\theta(t_1) - \theta(0))\\
& \nonumber = \frac{1}{2} (1-\theta(0)) t^2.
\end{align}

For the component $T^{(2)*}T^{(2)}$ the first integral is equal to:
\begin{align}
& \nonumber U^{2,2}_1(t) = \int\limits_0^t dt_1 \int\limits_0^{t_1} dt_2 \int\limits_0^{t} dt_3 \int\limits_0^{t_3} dt_4 \delta(t_1 - t_2) \delta(t_3 - t_4) \\
& \nonumber = \int\limits_0^t dt_1 \int\limits_0^{t_1} dt_2 \int\limits_0^{t} dt_3 \delta(t_1 - t_2) ( \theta(t_3) - \theta(0)) \\
& \nonumber = ( 1 - \theta(0)) t \int\limits_0^t dt_1 \int\limits_0^{t_1} dt_2 \delta(t_1 - t_2) \\
& \nonumber = ( 1 - \theta(0)) t \int\limits_0^t dt_1 ( \theta(t_1) - \theta(0)) \\
& \nonumber = ( 1 - \theta(0))^2 t^2.
\end{align}
Second integral:
\begin{align}
& \nonumber U^{2,2}_2(t) = \int\limits_0^t dt_1 \int\limits_0^{t_1} dt_2 \int\limits_0^{t} dt_3 \int\limits_0^{t_3} dt_4 \delta(t_1 - t_3) \delta(t_2 - t_4) \\
& \nonumber = \int\limits_0^t dt_1 \int\limits_0^{t_1} dt_2 \int\limits_0^{t} dt_3 (\theta(t_2) - \theta(t_2 - t_3)) \delta(t_1 - t_3)  \\
& \nonumber = \int\limits_0^t dt_1 \int\limits_0^{t_1} dt_2 (\theta(t_2) - \theta(t_2 - t_1)) (\theta(t_1) - \theta(t_1 - t)) \\
& \nonumber = \int\limits_0^t dt_1 \; t_1 (\theta(t_1) - \theta(-t_1)) (\theta(t_1) - \theta(t_1 - t)) \\
& \nonumber = \frac{1}{2} t^2.
\end{align}
Third integral:
\begin{align}
& \nonumber U^{2,2}_3(t) = \int\limits_0^t dt_1 \int\limits_0^{t_1} dt_2 \int\limits_0^{t} dt_3 \int\limits_0^{t_3} dt_4 \delta(t_1 - t_4) \delta(t_2 - t_3) \\
& \nonumber = \int\limits_0^t dt_1 \int\limits_0^{t_1} dt_2 \int\limits_0^{t} dt_3 (\theta(t_1) - \theta(t_1 - t_3)) \delta(t_2 - t_3) \\
& \nonumber = \int\limits_0^t dt_1 \int\limits_0^{t_1} dt_2 (\theta(t_1) - \theta(t_1 - t_2)) (\theta(t_2) - \theta(t_2 - t)) \\
& \nonumber = \int\limits_0^t dt_1 \; t_1 (\theta^2(t_1) - \theta(t_1)) = 0,
\end{align}
where we assumed $\theta(t)=1$ for all the integrals since $t>0$.

In the $d$-dimensional case second-order components of the transition probability form the following combination of 4-point correlation functions:
\begin{widetext}
\begin{align}
 & \mathbb{E}\Bigl[ \Bigl(\mathbf{w}(t_1) \cdot \mathbf{w}(t_2)\Bigr)\Bigl(\mathbf{w}(t_3) \cdot \mathbf{w}(t_4)\Bigr) + \Bigl(\mathbf{w}(t_1) \cdot \mathbf{w}(t_3)\Bigr)\Bigl(\mathbf{w}(t_2) \cdot \mathbf{w}(t_4)\Bigr) + \Bigl(\mathbf{w}(t_1) \cdot \mathbf{w}(t_4)\Bigr)\Bigl(\mathbf{w}(t_2) \cdot \mathbf{w}(t_3)\Bigr) \Bigr] \\
 & \nonumber = 3 \Bigl( \mathbb{E}[\mathbf{w}(t_1) \cdot \mathbf{w} (t_2)]\mathbb{E}[\mathbf{w}(t_3) \cdot \mathbf{w} (t_4)] + \mathbb{E}[\mathbf{w}(t_1) \cdot \mathbf{w} (t_3)]\mathbb{E}[\mathbf{w}(t_2) \cdot \mathbf{w} (t_4)] + \mathbb{E}[\mathbf{w}(t_1) \cdot \mathbf{w} (t_4)]\mathbb{E}[\mathbf{w}(t_2) \cdot \mathbf{w} (t_3)] \Bigr).
\end{align}
\end{widetext}
2-point correlation functions are equal to ones for 1-dimensional case, therefore the corresponding integrals can be calculated in the same manner as done above.

\section{Computations of the absolute masses}\label{numerics}

Here we stick to the scenario which considers spontaneous collapse as a sole source of the decay of neutral mesons (that is $\Gamma_{\mu}^{exp} = \Gamma_{\mu}^{CSL}$). We show here how we obtain the absolute masses from the experimental data given in~\cite{ParticleDataBook} by utilizing~(\ref{solintr}).  The procedure varies for each type of mesons. We start with $D$ and $B_d$ mesons. The authors of Ref.~\cite{ParticleDataBook} provide experimental values of the quantity $\Delta\Gamma / \Gamma$, namely
\begin{eqnarray}
 \frac{\Delta\Gamma}{\Gamma} =\left\lbrace\begin{array}{l}
\textrm{D-mesons: } \Biggl( 1.29 \left\lbrace\begin{array}{l} +0.14 \\
-0.18\end{array} \Biggr) \cdot 10^{-2} \right. ,\\
\textrm{B$_d$-mesons: } (0.1 \pm 1.0) \cdot 10^{-2}, \\
\end{array}\right.
\end{eqnarray}
where $\Delta\Gamma = \Gamma_L^{exp} - \Gamma_H^{exp}$ and $\Gamma = \frac{1}{2}(\Gamma_L^{exp} + \Gamma_H^{exp})$. Therefore we can easily obtain required decay rates for $D$ and $B_d$ mesons by dividing the quantity $\Delta\Gamma / \Gamma$ by 2
\begin{widetext}
\begin{eqnarray}
 \Biggl(\frac{\Gamma^{CSL}_L-\Gamma^{CSL}_H}{\Gamma^{CSL}_L+\Gamma^{CSL}_H}\Biggr)_{\mbox{D, B$_d$}} = \frac{1}{2} \frac{\Delta\Gamma}{\Gamma} =\left\lbrace\begin{array}{l}
\textrm{D-mesons: } 0.00645 \left\lbrace\begin{array}{l} +0.0007\\
-0.0009\end{array}\right.\\
\textrm{B$_d$-mesons: } 0.0005 \pm 0.005 \\
\end{array}\right.
\end{eqnarray}
Then we take into account mean lifetime of a meson $\tau = \frac{1}{\Gamma} = \frac{2}{\Gamma_L^{exp} + \Gamma_H^{exp}}$ and recover the decay constants for the light and heavy mass eigenstates
\begin{eqnarray}
& \nonumber \Gamma_L^{D, B_d} = &\frac{1}{2\tau} \Bigl(2 + \frac{\Delta\Gamma}{\Gamma}\Bigr) = \left\lbrace\begin{array}{l}
\textrm{D-mesons: } \Biggl( 2.4542 \left\lbrace\begin{array}{l} +0.006782\\
-0.007270 \end{array} \Biggr) \cdot 10^{12} \; \mbox{s$^{-1}$}\right. ,\\
\textrm{B$_d$-mesons: } (0.6582 \pm 0.001557 ) \cdot 10^{12} \; \mbox{s$^{-1}$}, \\
\end{array}\right. \\
& \nonumber \Gamma_H^{D, B_d} = &\frac{1}{2\tau} \Bigl(2 - \frac{\Delta\Gamma}{\Gamma}\Bigr) = \left\lbrace\begin{array}{l}
\textrm{D-mesons: } \Biggl( 2.4227 \left\lbrace\begin{array}{l} +0.011056\\
-0.010568 \end{array} \Biggr) \cdot 10^{12} \; \mbox{s$^{-1}$}\right. ,\\
\textrm{B$_d$-mesons: } (0.6576 \pm 0.005020 ) \cdot 10^{12} \; \mbox{s$^{-1}$}, \\
\end{array}\right.
\end{eqnarray}
where the errors are calculated up to the first order of Taylor series.

For $K$ and $B_s$ mesons the authors of Ref.~\cite{ParticleDataBook} provide the values of mean lifetimes of the corresponding mass eigenstates, $\tau_L$ for the light one (short-lived state as in the case of K-mesons) and $\tau_H$ for the heavy one (long-lived state as in the case of K-mesons)
\begin{eqnarray}
 & \tau_L = & \left\lbrace\begin{array}{l}
\textrm{K-mesons: } (0.8954 \pm 0.0004 ) \cdot 10^{-10} \; \mbox{s}, \\
\textrm{B$_s$-mesons: } (1.414 \pm 0.010 ) \cdot 10^{-12} \; \mbox{s}, \\
\end{array}\right.\\
 & \tau_H = & \left\lbrace\begin{array}{l}
\textrm{K-mesons: } (5.116 \pm 0.021 ) \cdot 10^{-8} \; \mbox{s}, \\
\textrm{B$_s$-mesons: } (1.624 \pm 0.014 ) \cdot 10^{-12} \; \mbox{s}.
\end{array}\right.
\end{eqnarray}
Using the definition of the decay constants of the mass eigenstate $\Gamma_{\mu} = \frac{1}{\tau_{\mu}}$ we obtain the following values
\begin{eqnarray}
 \Biggl(\frac{\Gamma^{CSL}_L-\Gamma^{CSL}_H}{\Gamma^{CSL}_L+\Gamma^{CSL}_H}\Biggr)_{\mbox{K, B$_s$}} = \frac{\frac{1}{\tau_L} - \frac{1}{\tau_H}}{\frac{1}{\tau_L} + \frac{1}{\tau_H}} = \left\lbrace\begin{array}{l}
\textrm{K-mesons: } 0.996506 \pm (1.2760 \cdot 10^{-5}), \\
\textrm{B$_s$-mesons: } 0.069124 \pm (7.7058 \cdot 10^{-4}), \\
\end{array}\right.
\end{eqnarray}
\end{widetext}
where the errors are calculated up to the first order of Taylor series. The decay constants for the mass eigenstates can be recovered by inverting the mean lifetimes
\begin{eqnarray}
 & \nonumber \Gamma_L^{K, B_s} = & \left\lbrace\begin{array}{l}
\textrm{K-mesons: } (1.1168 \pm 0.0005 ) \cdot 10^{10} \; \mbox{s$^{-1}$}, \\
\textrm{B$_s$-mesons: } (7.0721 \pm 0.010 ) \cdot 10^{11} \; \mbox{s$^{-1}$}, \\
\end{array}\right.\\
 & \nonumber \Gamma_H^{K, B_s} = & \left\lbrace\begin{array}{l}
\textrm{K-mesons: } (1.9547 \pm 0.0500 ) \cdot 10^{7} \; \mbox{s$^{-1}$}, \\
\textrm{B$_s$-mesons: } (6.1576 \pm 0.0531 ) \cdot 10^{11} \; \mbox{s$^{-1}$},
\end{array}\right.
\end{eqnarray}
where the errors are calculated up to the first order of Taylor series.

\end{document}